\documentclass{article}

\usepackage{geometry}

\usepackage[utf8]{inputenc}
\usepackage{amsfonts}
\usepackage{amsmath}
\usepackage{amssymb}
\usepackage{amsthm}
\usepackage{mathtools}
\usepackage{tikz}
\usepackage{graphicx}
\usepackage{braket}
\usepackage{tikz-cd}
\usepackage{breqn}
\usepackage{hyperref}
\usepackage{booktabs}
\hypersetup{
    colorlinks=false,
    linkcolor=blue,
    filecolor=magenta,      
    urlcolor=cyan,
    pdftitle={Overleaf Example},
    pdfpagemode=FullScreen,
}
\usepackage{blkarray}
\usepackage{color}
\usepackage{comment}
\usepackage{enumerate}
\usepackage{upgreek}

\usepackage[ruled,vlined]{algorithm2e}
\usepackage{enumerate}
\usepackage{amsfonts}                  
\usepackage{enumitem}
\usepackage[normalem]{ulem}
\usepackage{blkarray}
\usepackage{multirow}
\usepackage{blkarray, bigstrut}
\usepackage{xparse}
\usepackage{mhchem}
\usepackage[toc,page]{appendix}

\theoremstyle{definition}

\newtheorem{theorem}{Theorem}

\newtheorem{assumption}{Assumption}

\usepackage{natbib}

\newcommand\cmnt[2]{\;
{\textcolor{red}{[{\em #1 --- #2}] \;}
}}
\newcommand\benji[1]{\cmnt{#1}{Benji}}

\newcommand\jia[1]{\cmnt{#1}{Jia}}

\newcommand{\escore}[1]{\frac{\exp(#1)}{1 + \exp(#1)}}

\makeatletter
\DeclareRobustCommand{\pdot}{\mathbin{\mathpalette\pdot@\relax}}
\newcommand{\pdot@}[2]{%
  \ooalign{%
    $\m@th#1\circ$\cr
    \hidewidth$\m@th#1\cdot$\hidewidth\cr
  }%
}
\makeatother

\title{Quantifying the Uncertainty of Imputed Demographic Disparity Estimates: The Dual-Bootstrap}
\author{%
    Benjamin Lu\thanks{University of California, Berkeley},
    Jia Wan\thanks{Massachusetts Institute of Technology},
    Derek Ouyang\thanks{Stanford University},\\
    Jacob Goldin\thanks{University of Chicago and NBER},
    Daniel E. Ho\thanks{Stanford University}
}

\begin{document}
\maketitle

\begin{abstract}
    Measuring average differences in an outcome across racial or ethnic groups is a crucial first step for equity assessments, but researchers often lack access to data on individuals’ races and ethnicities to calculate them. A common solution is to impute the missing race or ethnicity labels using proxies, then use those imputations to estimate the disparity. Conventional standard errors mischaracterize the resulting estimate's uncertainty because they treat the imputation model as given and fixed, instead of as an unknown object that must be estimated with uncertainty. We propose a dual-bootstrap approach that explicitly accounts for measurement uncertainty and thus enables more accurate statistical inference, which we demonstrate via simulation. In addition, we adapt our approach to the commonly used Bayesian Improved Surname Geocoding (BISG) imputation algorithm, where direct bootstrapping is infeasible because the underlying Census Bureau data are unavailable. In simulations, we find that measurement uncertainty is generally insignificant for BISG except in particular circumstances; bias, not variance, is likely the predominant source of error. We apply our method to quantify the uncertainty of prevalence estimates of common health conditions by race using data from the American Family Cohort.
\end{abstract}

\section{Introduction}
\label{sec:intro}
Racial and ethnic disparities are a common focus of academic study, policymaking, and advocacy efforts across many domains, including criminal justice \citep{gelman2007analysis, berdejo2018criminalizing}, health care \citep{azin2020racial, mackey2021racial}, technology \citep{buolamwini2018gender, koenecke2020racial}, and taxation \citep{brown2022whiteness, avenancio2022assessment}.\footnote{For brevity, we use ``race'' to refer to both race and ethnicity throughout the remainder of this paper.}

Such disparities are straightforward to compute---if individual-level demographic and outcome data are available. In many settings where the measurement of racial disparities is of interest, however, race data are missing or otherwise inaccessible. For example, Regulation B of the Equal Credit Opportunity Act prohibits creditors from discriminating against an applicant on the basis of race. But monitoring and enforcing this prohibition is complicated by the fact that the very same laws also prohibit creditors from inquiring about an applicant's race at all.

Some researchers work around this problem by imputing individuals' races based on observable proxy features, then using those imputations to estimate the racial disparity. One of the most common imputation methods is Bayesian Improved Surname Geocoding (BISG), which imputes an individual's race based on their surname and geolocation \citep{elliott2009using}. Recent work has also investigated the potential power of machine learning for this task \citep{cheng2023, xue2019comparison, kim2018riddle}.

But imputations are estimates, not oracles. Like any other statistic, each imputation is the output of an estimator fit on data, with its own bias and variance. This statistical uncertainty could affect the quality of the downstream racial disparity estimate. Many studies that use imputations ignore this potential error propagation, instead treating the imputation model as known with certainty \citep[e.g.,][]{brown2016using, zhang2018, yee2022implications}. Doing so can imperil the reliability of the final estimate in different ways \citep[e.g.,][adjusting for imputation bias]{labgold2021}.

This paper examines one aspect of the problem: the effect of measurement uncertainty on statistical inference. It is standard practice in academic research to report the confidence interval or standard error associated with a racial disparity estimate. But typical confidence intervals and standard errors reflect only classical \textit{sampling uncertainty}---i.e., uncertainty arising from the fact that the disparity estimate is based on only a sample of the broader population of interest. They do not reflect the \textit{measurement uncertainty} that arises from estimating the race probability model and thus risk mischaracterizing the degree of confidence in the disparity estimate. 

We make three contributions to the study of this issue. First, we offer a ``dual-bootstrap'' procedure that incorporates both sampling and measurement uncertainty and thus offers more accurate statistical inference. We prove that our procedure is consistent for some race probability models under standard regularity conditions.

Second, we adapt our procedure to the special case of BISG, where Census Bureau-imposed constraints on data availability raise particular challenges. The Census Bureau does not disclose the individual-level survey responses on which its popularly used American Community Survey race-by-geolocation estimates are based. This prevents researchers from directly applying the general dual-bootstrap algorithm. We propose one way to nonetheless approximate the measurement uncertainty of BISG race probability estimates using other information provided by the Census Bureau.

Third, we apply our approach to simulated and real data to investigate how much measurement uncertainty contributes to the final disparity estimate's standard error. Our findings suggest that, in general, the uncertainty of BISG imputations only negligibly increases standard errors because BISG is a relatively inflexible model based on large-scale data (i.e., full-scale census records); bias, not variance, is likely the predominant type of error in BISG. But we do find some exceptions: BISG measurement uncertainty, and the way it is estimated, can substantially affect the final inference when studying particular demographic or geographic groups. We also show that when race probability models more flexible than BISG are employed, properly accounting for measurement uncertainty can substantially affect the widths of resulting confidence intervals. We illustrate these findings through an analysis of racial disparities in common health outcomes in the American Family Cohort, a dataset containing the electronic health records of primary care visits by patients in the United States. Our method has also been applied in a recent working paper studying racial disparities in tax audit rates \citep{elzayn2023measuring}.

\section{Related Work}
\label{sec:related_work}

To our knowledge, the role of measurement uncertainty in the specific context of race imputation has not been thoroughly studied. As mentioned in Section \ref{sec:intro}, many studies where race is imputed simply ignore it. One exception is a concurrent working paper by \citet{derby2024}, who propose a fully Bayesian approach where a prior distribution for the conditional race probabilities is assumed, then updated based on reported Census Bureau estimates to obtain a posterior distribution from which conditional race probabilities are sampled.\footnote{\citet{imai2022} similarly impose a prior, but they focus on how doing so improves the accuracy of race predictions, not on how it can more accurately quantify the uncertainty of downstream estimates.} Our proposal for BISG is similar in spirit. It can be viewed as the frequentist analog---but with the distinct advantage of constructing a sampling distribution of the conditional race probability estimates based on the uncertainty that the Census Bureau actually reports for those estimates, instead of a purely assumed prior model. This fidelity comes at some cost to flexibility; see Section \ref{sec:bisg} for discussion.

This paper draws from a rich body of research on missing data, survey design, and causal inference. Especially relevant are two strands of work, on inverse propensity weighting (IPW) and $Z$-estimation. The disparity estimator we consider, analyzed by \citet{chen2019fairness} and described in Section \ref{sec:setup} below, weights individuals by their estimated probability of being of a given race. It is thus very similar in form to H\'{a}jek IPW estimators, which have been extensively studied by, for example, \citet{miratrix2018} and \citet{matsouaka2022overlap}. And, following some prior work on the properties of IPW estimators in the context of causal inference \citep{reifeis2022variance, shu2021variance}, we rely on $Z$-estimator theory to establish the asymptotic properties of our proposed method \citep{kosorok2008introduction, stefanski2002calculus}.

We distinguish our subject of investigation from several other important but distinct areas of study. First, we focus solely on the variability of the imputed disparity estimator, as typically reflected in metrics like the standard error or confidence intervals. Prior work has examined identification and bias properties of the specific imputed disparity estimator on which we focus \citep{chen2019fairness, kallus2022assessing, elzayn2023measuring}. Others have examined the accuracy and bias of specific race imputation models that often underlie imputed disparity estimators. For example, as mentioned above, \citet{imai2022} propose ways of improving the accuracy of BISG by accounting for the possible migration of racial minorities to geographic areas where none resided prior to the latest census count. These issues are largely orthogonal to the challenge of accurately characterizing an imputation-based estimator's variability. In our theory and simulations, we assume that these issues have been favorably resolved.

Second, our work is distinct from multiple imputation, at least in its classical formulation. In the typical setting amenable to multiple imputation, race is observed as a categorical variable for a subset of the data to be analyzed, and the researcher seeks to impute the categorical race variable for the remaining subset of the data where it is missing; \citet{fong2021machine} discuss some challenges of proper statistical inference in that setting that are similar to the ones we address here. But the setting we consider (described in Section \ref{sec:setup}) is one where race is completely unobserved for the data to be analyzed, and the researcher seeks to estimate each unit's real-valued probability of being a given race, not the unit's actually realized race. Our problem setting is thus more closely related to that of measurement error models \citep[e.g.,][]{fuller2009} and two- or split-sample instrumental variables \citep[e.g.,][]{angrist1992effect, angrist1995split}, with particular focus on uncertainty quantification for general, nonlinear measurement models. Nonetheless, some conceptual similarities to multiple imputation can be drawn. Perhaps the most salient connection is to the concept of ``proper'' multiple imputation, defined by \citet{rubin1987}. As \citet{murray2018} summarizes it, multiple imputation generally yields valid inference only if, among other conditions, the uncertainty of the imputation model itself is accounted for. This same concept underpins our work.

\section{Setup and Notation}
\label{sec:setup}

Consider two datasets: a training dataset $\mathcal{T} \equiv \{Z_i, A_i\}_{i = 1}^{n_{\mathcal{T}}}$ and a primary dataset $\mathcal{P} \equiv \{Z_j, Y_j\}_{j = 1}^{n_{\mathcal{P}}}$, where $Y$ denotes the outcome, $A$ is a binary indicator of race, $Z$ denotes observable proxies of race, and $n_{\mathcal{T}}$ and $n_{\mathcal{P}}$ are the number of units in the training and primary datasets, respectively. The training dataset is drawn i.i.d. from some population $\mathbb{T}$, and the primary dataset is drawn i.i.d. from a potentially different population $\mathbb{P}$. Our estimand is the racial disparity in outcomes in the primary population $\mathbb{P}$:
\[
\delta \equiv \mathbb{E}_{\mathbb{P}}\left[Y \mid A = 1\right] - \mathbb{E}_{\mathbb{P}}\left[Y \mid A = 0\right].
\]

If $A$ were observed in the primary dataset, estimation and inference of $\delta$ would be straightforward. But it is not---so we impute $A$ using $Z$ based on some model class $\mathcal{F}_A$ instead. Specifically, we fit a model $f \in \mathcal{F}_A$ of $A$ on $Z$ using the training dataset, where $(Z, A)$ is jointly observed. We then use that model to estimate the race probability of each unit in the primary dataset: $\widehat{\Pr}_{\mathbb{P}}\left(A = 1 \mid Z = Z_j\right) \equiv f(Z_j)$. Finally, we estimate the racial disparity by the probabilistic weighting estimator that \citet{chen2019fairness} propose:
\[
\hat{\delta} \equiv \frac{\sum_{j = 1}^{n_{\mathcal{P}}}\widehat{\Pr}_{\mathbb{P}}\left(A = 1 \mid Z = Z_j\right)Y_j}{\sum_{j = 1}^{n_{\mathcal{P}}}\widehat{\Pr}_{\mathbb{P}}\left(A = 1 \mid Z = Z_j\right)} - \frac{\sum_{j = 1}^{n_{\mathcal{P}}}\widehat{\Pr}_{\mathbb{P}}\left(A = 0 \mid Z = Z_j\right)Y_j}{\sum_{j = 1}^{n_{\mathcal{P}}}\widehat{\Pr}_{\mathbb{P}}\left(A = 0 \mid Z = Z_j\right)}.
\]

Other imputation-based disparity estimators have been used or analyzed in past work. For example, \cite{chen2019fairness} discuss a thresholding estimator that estimates the mean outcome in each race by classifying individuals' races instead of using soft probabilities. And \citet{elzayn2023measuring} consider the slope coefficient in a linear regression of the outcomes on the estimated race probabilities, which they show can in conjunction with $\hat{\delta}$ bound the true disparity. But we focus on $\hat{\delta}$ because it is a commonly used estimator with favorable statistical properties \citep{chen2019fairness, mccaffrey2008power}. We briefly outline how our framework might extend to the linear disparity estimator of \citet{elzayn2023measuring} in Appendix \ref{appendix:other_z_est}, but we defer a detailed examination of this and other extensions to future work.

We invoke standard assumptions so that the estimator $\hat{\delta}$ is consistent for $\delta$. First, we assume that the probability model $\widehat{\Pr}_{\mathbb{P}}\left(A = 1 \mid Z\right)$ is correctly specified. Since the model is fit on the training dataset $\mathcal{T}$ but used to characterize the primary population $\mathbb{P}$, this assumption typically also implies that the conditional distribution of $A$ given $Z$ is the same in $\mathbb{T}$ and $\mathbb{P}$ everywhere $Z$ has positive density in $\mathbb{P}$. Second, we assume that $\mathbb{E}_{\mathbb{P}}\left[\text{Cov}_{\mathbb{P}}\left(A, Y \mid Z\right)\right] = 0$; \citet{chen2019fairness} show that this condition is sufficient for $\hat{\delta}$ to be consistent when the true probabilities are given. Since our focus is inference, not estimation, we take these assumptions for granted and refer interested readers to past work on the consistency of $\hat{\delta}$.

\section{The Dual-Bootstrap}

\begin{figure}[h]
    \centering
    \includegraphics[width = 0.99\linewidth]{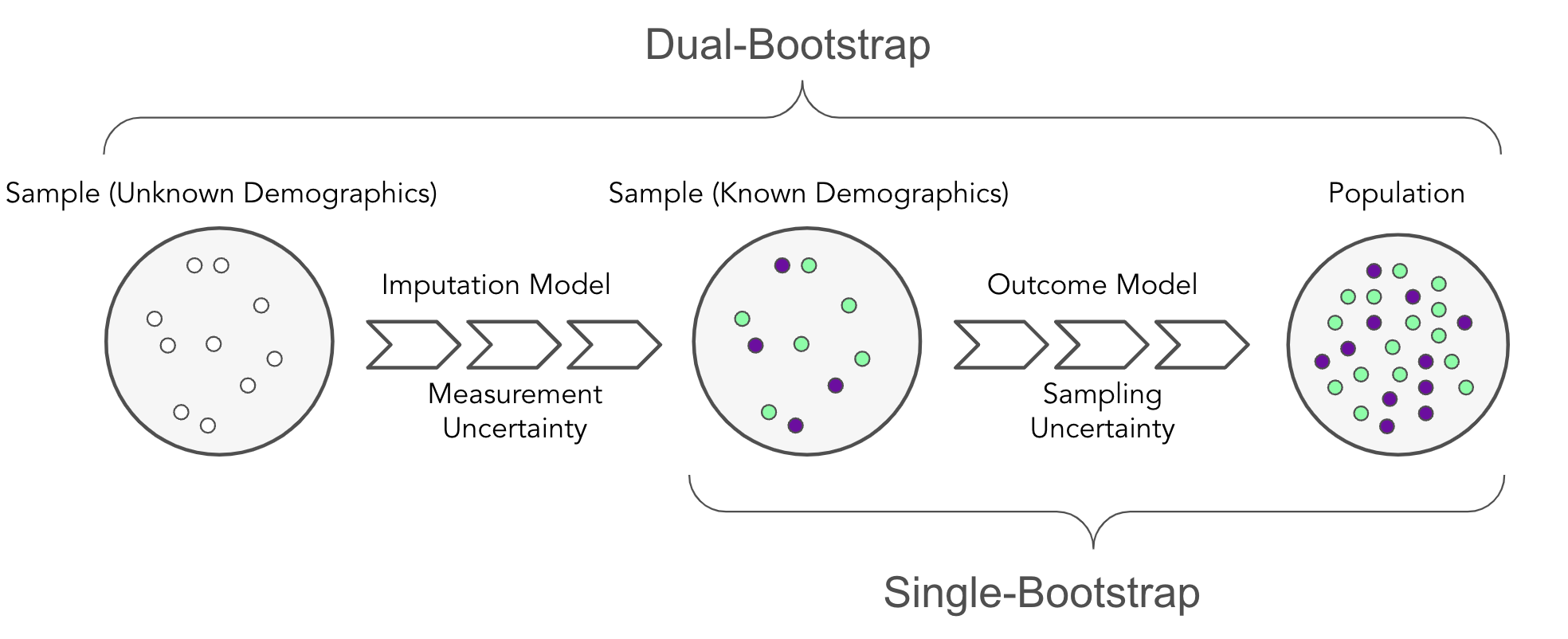}
    \caption{Illustration of the uncertainties captured by the dual-bootstrap, compared to those captured by the single-bootstrap.}    \label{fig:single_dual_bootstrap_diagram}
\end{figure}

We propose a ``dual-bootstrap'' procedure to enable proper inference of $\hat{\delta}$ that accounts for both sampling and measurement uncertainty, as Figure \ref{fig:single_dual_bootstrap_diagram} illustrates. We first state the procedure in general terms, then investigate via simulation the effects of measurement uncertainty and the dual-bootstrap's ability to account for it.

\subsection{General Procedure}
\label{sec:gen_pro}

Algorithm \ref{alg:dual_bootstrap} states the dual-bootstrap procedure. It frames the desired output as a confidence interval estimated via the percentile bootstrap, but other uncertainty metrics can be estimated too. As Algorithm \ref{alg:dual_bootstrap} shows, the dual-bootstrap is straightforward: We simply resample with replacement both the training and the primary datasets, then refit the race probability model on the resampled training dataset and apply it to estimate the racial disparity in the resampled primary dataset. The algorithm here calls for simple resampling with replacement, but other, more complex forms of resampling may be appropriate---for example, if the data are clustered \citep{owen2007pigeonhole, derby2024}.

The key contribution of the dual-bootstrap stems from its resampling of the training dataset and refitting of the race probability model. Doing so accounts for the uncertainty of the race probability estimates themselves. This uncertainty is then propagated downstream to the bootstrap statistic $\hat{\delta}^{\ast b}$. As discussed above, some prior work has ignored this measurement uncertainty entirely, instead treating $\widehat{\Pr}_{\mathbb{P}}\left(A = 1 \mid Z = Z_j\right)$ as true. This corresponds to skipping the first two lines of the for-loop in Algorithm \ref{alg:dual_bootstrap}.

We prove in Appendix \ref{appendix:corrected_proof} that $\hat{\delta}$ and its dual-bootstrap analogs $\hat{\delta}^{\ast b}$ are asymptotically normal when the race probabilities are estimated via logistic regression and general regularity conditions hold. We limit the proof to logistic regression because doing so allows us to frame $\hat{\delta}$ as a $Z$-estimator---broadly, any estimator that can be expressed as the approximate zero of a data-dependent function---to which standard theoretical results can apply. The proof strategy likely applies readily to other race probability models that fall within the $Z$-estimation framework, albeit possibly with slight modifications to the regularity conditions. We leave such extensions to future work. It is less clear what theoretical properties the dual-bootstrap of $\hat{\delta}$ has when the race probability model does not fall within the $Z$-estimation framework. A closer examination of this issue might prove fruitful.

The $Z$-estimation theory that we apply to prove asymptotic normality also provides a closed-form expression for the variance of the limiting distribution, but this is mostly of theoretical interest. In practice, deriving the closed-form expression usually has little utility. Statistical software like the \texttt{geex} package in \texttt{R} can compute the empirical variance estimator using numerical routines, without requiring analytic derivations \citep{geex}. When the race probability model falls within the $Z$-estimation framework, using such numerical solvers can often require much less computational power than the dual-bootstrap.

\begin{algorithm}[hbt!]
\DontPrintSemicolon
\caption{Dual-Bootstrap}\label{alg:dual_bootstrap}
\KwData{Training Dataset $\mathcal{T} = \{Z_i, A_i\}_{i = 1}^{n_{\mathcal{T}}}$, Primary Dataset $\mathcal{P} = \{Z_j, Y_j\}_{j = 1}^{n_{\mathcal{P}}}$, Model Class $\mathcal{F}_{A}$, Number of Bootstrap Draws $B \in \mathbb{N}$, Level $\alpha \in [0, 1]$}
\KwResult{Confidence interval for the demographic disparity estimate $\hat{\delta}$}
\For{$b$ in range $B$}{
    Resample $\mathcal{T}^{\ast b}$ by sampling with replacement from $\mathcal{T}$\;
    Fit $\widehat{\Pr}^{\ast b}_{\mathbb{P}}\left(A = 1 \mid Z\right) \in \mathcal{F}_A$ on $\mathcal{T}^{\ast b}$ \;
    Resample $\mathcal{P}^{\ast b}$ by sampling with replacement from $\mathcal{P}$ \;
    Compute 
    \[
    \hat{\delta}^{\ast b} \equiv \frac{\sum_{j = 1}^{n_{\mathcal{P}}}\widehat{\Pr}^{\ast b}_{\mathbb{P}}\left(A = 1 \mid Z = Z^{\ast b}_j\right)Y^{\ast b}_j}{\sum_{j = 1}^{n_{\mathcal{P}}}\widehat{\Pr}^{\ast b}_{\mathbb{P}}\left(A = 1 \mid Z = Z^{\ast b}_j\right)} - \frac{\sum_{j = 1}^{n_{\mathcal{P}}}\widehat{\Pr}^{\ast b}_{\mathbb{P}}\left(A = 0 \mid Z = Z^{\ast b}_j\right)Y^{\ast b}_j}{\sum_{j = 1}^{n_{\mathcal{P}}}\widehat{\Pr}^{\ast b}_{\mathbb{P}}\left(A = 0 \mid Z = Z^{\ast b}_j\right)}
    \]
}
Output $(1-\alpha)$-level percentile bootstrap confidence interval
\[\left( \hat{\delta}_B^{(\alpha/2)}, \hat{\delta}_B^{(1-\alpha/2)} \right)\]
where $\hat{\delta}_B^{(\alpha)}$ is the empirical $\alpha$-percentile of the $\hat{\delta}^{\ast b}$
\end{algorithm}

\subsection{Simulations}
\label{sec:ml_sim}

We demonstrate empirically that the dual-bootstrap more accurately accounts for the overall uncertainty of imputed disparity estimates. We do so through a simple simulation in which both the training and primary populations follow the same data-generating process:
\begin{itemize}
    \item A single proxy is drawn i.i.d. from a standard normal: $Z \sim \mathcal{N}(0, 1)$.
    \item Race is drawn i.i.d. from a Bernoulli distribution with probability logistic in $Z$: $A \mid Z \sim \text{Bern}\left[\exp(Z) / \{\exp(Z) + 1\}\right]$.
    \item The outcome $Y$ is i.i.d. normal and linear in $Z$: $Y \mid Z \sim \mathcal{N}(5Z, 9)$.
\end{itemize}

In each simulation repetition, we draw $(Z, A)$ tuples as the training dataset $\mathcal{T}$ and $(Z, Y)$ tuples as the primary dataset $\mathcal{P}$. We fit a logistic regression of $A$ on $Z$ with $\mathcal{T}$, then apply it to $\mathcal{P}$ to obtain our point estimate $\hat{\delta}$. We then apply the dual-bootstrap with 2,000 bootstrap iterations to estimate a 95\% confidence interval for $\delta$. For comparison, we also estimate what we call the ``single-bootstrap'' standard 95\% confidence interval, in which the race probability estimates are treated as given and only $\mathcal{P}$ is resampled. We also estimate a 95\% confidence interval based on the empirical variance estimator implied by $Z$-estimation theory using the \texttt{geex} package.

Table \ref{tab:ml_sim} reports the resulting coverage rates of the three types of confidence intervals over 500 simulation repetitions for various sample sizes of $\mathcal{T}$ and $\mathcal{P}$. We note four trends. First, the coverage rate of the single-bootstrap is worst when $\mathcal{T}$ is small and best, though still inadequate, when it is large. This reflects the effect of measurement uncertainty on the variance of the ultimate disparity estimator in this specific simulation setup; as $\mathcal{T}$ increases, the variability of the imputations decreases and becomes less influential. Second, for any size of $\mathcal{T}$, the single-bootstrap's coverage rate decreases as $\mathcal{P}$ increases. This suggests that the importance of measurement uncertainty can amplify with the number of imputations required. Third, the dual-bootstrap provides better coverage than the single-bootstrap, but it requires sufficiently large sample sizes to achieve the desired 95\% rate. This reflects the fact that the theoretical properties of the bootstrap take effect asymptotically, as we discuss in Appendix \ref{appendix:corrected_proof}. Finally, the empirical variance estimator and the dual-bootstrap behave nearly identically; this is consistent with the theoretical results of Appendix \ref{appendix:corrected_proof}.

\begin{table}
\centerline{
\begin{tabular}[t]{@{}cccccccc@{}}
\toprule
& & \multicolumn{3}{c}{Coverage Rate} & \multicolumn{3}{c}{Interval Width} \\
\cmidrule(l{3pt}r{3pt}){3-5} \cmidrule(l{3pt}r{3pt}){6-8} $\left|\mathcal{T}\right|$ & $\left|\mathcal{P}\right|$ & Dual-Bootstrap & Single-Bootstrap & Empirical & Dual-Bootstrap & Single-Bootstrap & Empirical\\
\midrule
100 & 100 & 0.91 & 0.81 & 0.92 & 3.8 & 2.2 & 3.7\\
100 & 1000 & 0.94 & 0.66 & 0.95 & 3.1 & 0.7 & 3.0\\
100 & 5000 & 0.94 & 0.54 & 0.94 & 3.0 & 0.3 & 3.0\\
1000 & 100 & 0.89 & 0.87 & 0.90 & 2.4 & 2.2 & 2.4\\
1000 & 1000 & 0.93 & 0.80 & 0.93 & 1.2 & 0.7 & 1.2\\
1000 & 5000 & 0.97 & 0.67 & 0.97 & 1.0 & 0.3 & 1.0\\
5000 & 100 & 0.87 & 0.86 & 0.87 & 2.2 & 2.2 & 2.2\\
5000 & 1000 & 0.88 & 0.84 & 0.88 & 0.8 & 0.7 & 0.8\\
5000 & 5000 & 0.94 & 0.83 & 0.94 & 0.5 & 0.3 & 0.5\\
\bottomrule
\end{tabular}}
\caption{Coverage rates and widths of 95\% confidence intervals estimated using the dual-bootstrap, the single-bootstrap, and the empirical variance estimator for varying sample sizes of $\mathcal{T}$ and $\mathcal{P}$. \label{tab:ml_sim}}
\end{table}

\section{Special Case: BISG}
\label{sec:bisg}

In this section, we adapt the dual-bootstrap to the BISG algorithm, which is commonly used in empirical applications as the race probability model. The key challenge to applying our dual-bootstrap approach in this setting is that the Census Bureau does not generally make publicly available the training dataset $\mathcal{T}$ on which the BISG prior probabilities are based. We suggest one way to overcome this constraint while still upholding the fundamental principle animating the generic dual-bootstrap procedure of Section \ref{sec:gen_pro}. We then apply our method to assess how prominent the uncertainty of BISG imputations are in practice. We conclude that, with some notable exceptions, the variability of BISG imputations is generally negligible in practice; bias, not variance, is likely the primary source of error in BISG.

\subsection{BISG-Specific Procedure}
\label{sec:bisg_procedure}

BISG imputes race by naively applying Bayes' Theorem to Census Bureau estimates of the racial composition of people by surname and geolocation. In this context, $A$ is typically categorical instead of binary, containing all race categories defined by the Census Bureau; $Z = (S, G)$, where $S$ is a categorical variable denoting the individual's surname and $G$ is a categorical variable denoting the individual's geolocation; and the race probability model is
\[
\widehat{\Pr}_{\mathbb{P}}\left(A = a' \mid S = s, G = g\right) \equiv \frac{\widehat{\Pr}_{\mathbb{P}}\left(A = a' \mid G = g\right)\widehat{\Pr}_{\mathbb{P}}\left(S = s \mid A = a'\right)}{\sum_{a}\widehat{\Pr}_{\mathbb{P}}\left(A = a \mid G = g\right)\widehat{\Pr}_{\mathbb{P}}\left(S = s \mid A = a\right)},
\]
where the prior probabilities on the right are parameter estimates based on Census Bureau surveys. Specifically, researchers commonly use as $\widehat{\Pr}_{\mathbb{P}}\left(A \mid G\right)$ the Census Bureau's American Community Survey (ACS) estimates of the number of people of each race residing in each geolocation. And they compute $\widehat{\Pr}_{\mathbb{P}}\left(S \mid A\right)$ based on the Census Bureau's 2010 table of frequently occurring surnames. For BISG, then, the training dataset $\mathcal{T}$ is the microdata---i.e., individual-level survey responses---that the Census Bureau collects to generate the ACS and surname estimates. The standard dual-bootstrap procedure outlined in Algorithm \ref{alg:dual_bootstrap} thus calls for the analyst to resample the microdata with replacement and recompute the racial composition of each geolocation and surname.\footnote{The surname table is a raw tabulation of data from the decennial census, which covers the entire population of the United States. Thus, depending on the population for which race-specific outcomes are to be estimated, resampling the data that produced the surname table might be unnecessary. Nonetheless, we outline our procedure to include resampling for two reasons. First, it could be appropriate to do so, depending on the estimand. Second, race-by-name probabilities are sometimes sourced from data that are properly characterized as a sample, rather than the entire population. For example, \citet{imai2022} use voting records from a handful of states to estimate the race-by-name probabilities. In such cases, resampling would likely be appropriate.}

The challenge, however, is that $\mathcal{T}$ is inaccessible. The Census Bureau generally does not publish microdata for privacy reasons. This is not an issue for the microdata on which the surname-race probabilities are based: Since the surname table is just a raw tabulation, we can still essentially reconstruct the microdata that produced it and resample from it.\footnote{Such a reconstruction is necessarily imprecise since the surname table aggregates all surnames held by fewer than 100 individuals into a single ``Other'' category. Our reconstruction of the microdata can never recover these surnames. We leave a detailed examination of the significance of this issue to future work.} But the same is not true of the microdata on which the ACS race-by-geolocation estimates are based. The ACS estimates of the racial composition of geographic areas are not raw tabulations of survey microdata; rather, they are produced by re-weighting the microdata to adjust for factors like probability of selection in an unknown and presumably complex way.\footnote{Some ACS microdata are available, but only at levels of geographic granularity that are too low to be useful in most applications. Such microdata are also incomplete---they consist of only about two-thirds of the records used to produce the ACS estimates---and thus might not be any more amenable to direct resampling.} Thus, a solution for the race-by-geolocation probabilities is required.

The key intuition behind the solution we propose is that we seek to resample the microdata $\mathcal{T}$ and recompute the prior $\widehat{\Pr}_{\mathbb{P}}\left(A \mid G\right)$ only as a means of, essentially, drawing from the approximate sampling distribution of $\widehat{\Pr}_{\mathbb{P}}\left(A \mid G\right)$. If we can approximate the sampling distribution of $\widehat{\Pr}_{\mathbb{P}}\left(A \mid G\right)$ in some other way, then we can just draw from it directly---no resampling or model-refitting needed. Fortunately, the Census Bureau suggests and endorses a way of estimating key parameters of the sampling distribution even without microdata. We outline the approach below, with further details available in Appendix \ref{appendix:bisg}.

As an initial matter, we center the sampling distribution of $\widehat{\Pr}_{\mathbb{P}}\left(A \mid G\right)$ at the published ACS estimate, which we denote by $\hat{\mu}_G$. Then, to estimate the covariance of this sampling distribution, which we denote by $\hat{\Sigma}_G$, we use the publicly available ACS variance replicates. These are 80 ``pseudo-estimates'' of the racial composition of each geolocation, which we denote by $\hat{\mu}^{\dagger r}_G$ for $r = 1, \ldots, 80$. The Census Bureau uses them to estimate variances via the successive differences replication (SDR) method. The variance replicates are not bootstrap statistics; they have ``no other use [beyond calculating SDR variances] and no independent meaning'' \citep{SDRCensus}. So we cannot directly apply the dual-bootstrap algorithm to them. Instead, we use the variance replicates to estimate the covariance of the race-by-geolocation probability estimates based on the formula prescribed by the Census Bureau:
\[
\hat{\Sigma}_G \equiv \frac{4}{80}\sum_{r = 1}^{80}\left(\hat{\mu}^{\dagger r}_G - \hat{\mu}_G\right)\left(\hat{\mu}^{\dagger r}_G - \hat{\mu}_G\right)^{\intercal}.
\]
When ACS estimates that there are zero people of a given race in a geographic area, all associated variance replicates are zero. In such ``zero-count'' cases, we follow the Census Bureau's recommendation not to use the above formula; instead, we assume the estimate has zero covariance with the other estimates and essentially derive the variance from the Census Bureau's estimated margin of error \citep{SDRCensus}. Appendix \ref{sec:zero_counts} describes the procedure in more detail. As we discuss in Section \ref{sec:bisg_sim}, the choice to account for uncertainty in zero-count geolocations can be influential in specific circumstances.

Finally, we assume that the sampling distribution of $\widehat{\Pr}_{\mathbb{P}}\left(A \mid G\right)$ is normal, so the parameter estimates $(\hat{\mu}_G, \hat{\Sigma}_G)$ fully specify the sampling distribution as $\mathcal{N}(\hat{\mu}_G, \hat{\Sigma}_G)$. We leave generalizations of the form of the sampling distribution to future work. See Appendix \ref{sec:impermissible} for more discussion.

With the sampling distribution of $\widehat{\Pr}_{\mathbb{P}}\left(A \mid G\right)$ in hand, our modified dual-bootstrap routine can be executed. Algorithm \ref{alg:bisg} states the modified implementation. The key distinction from Algorithm \ref{alg:dual_bootstrap} is that the bootstrap race-by-geolocation probability estimate $\widehat{\Pr}^{\ast b}_{\mathbb{P}}\left(A \mid G\right)$ is computed by drawing directly from the sampling distribution $\mathcal{N}(\hat{\mu}_G, \hat{\Sigma}_G)$ instead of by refitting on resampled microdata. As with Algorithm \ref{alg:dual_bootstrap}, the resampling steps in this algorithm use simple resampling with replacement, but more complex forms of resampling may be appropriate \citep{owen2007pigeonhole, derby2024}.

\begin{algorithm}[hbt!]
\DontPrintSemicolon
\caption{BISG Dual-Bootstrap}\label{alg:bisg}
\KwData{ACS Estimate $\hat{\mu}$, ACS Covariance Matrix Estimate $\hat{\Sigma}$, Surname Table $\mathcal{S}$, Primary Dataset $\mathcal{P} = \{S_j, G_j, Y_j\}_{j = 1}^{n_{\mathcal{P}}}$, Number of Bootstrap Draws $B \in \mathbb{N}$, Level $\alpha \in [0, 1]$, Race Groups $a'$ and $a''$}
\KwResult{Confidence interval for the demographic disparity estimate $\hat{\delta}$}
\For{$b$ in range $B$}
{
    Resample $\mathcal{S}^{\ast b}$ by sampling with replacement from $\mathcal{S}$ \tcp*{optional; see Footnote 3}
    \For{s in $\{S_j\}_{j = 1}^{n_{\mathcal{P}}}$, a in \text{supp}(A)}{
    Compute $\widehat{\Pr}^{\ast b}_{\mathbb{P}}\left(S = s \mid A = a\right)$ from $\mathcal{S}^{\ast b}$
    \tcp*{optional; see Footnote 3}}
    \For{g in $\{G_j\}_{j = 1}^{n_{\mathcal{P}}}$} {
    Sample $\widehat{\Pr}^{\ast b}_{\mathbb{P}}\left(A \mid G = g\right) \sim \mathcal{N}\left(\hat{\mu}_g, \hat{\Sigma}_g\right)$ \;
    }
    Resample $\mathcal{P}^{\ast b}$ by sampling with replacement from $\mathcal{P}$ \;
    \For{j in range $n_{\mathcal{P}}$}{
    Compute 
    \begin{align*}
    \widehat{\Pr}^{\ast b}_{\mathbb{P}}\left(A = a' \mid S = S^{\ast b}_j, G = G^{\ast b}_j\right) &\equiv \frac{\widehat{\Pr}^{\ast b}_{\mathbb{P}}\left(A = a' \mid G = G^{\ast b}_j\right)\widehat{\Pr}^{\ast b}_{\mathbb{P}}\left(S = S^{\ast b}_j \mid A = a'\right)}{\sum_{a}\widehat{\Pr}^{\ast b}_{\mathbb{P}}\left(A = a \mid G = G^{\ast b}_j\right)\widehat{\Pr}^{\ast b}_{\mathbb{P}}\left(S = S^{\ast b}_j \mid A = a\right)},\\
    \widehat{\Pr}^{\ast b}_{\mathbb{P}}\left(A = a'' \mid S = S^{\ast b}_j, G = G^{\ast b}_j\right) &\equiv \frac{\widehat{\Pr}^{\ast b}_{\mathbb{P}}\left(A = a'' \mid G = G^{\ast b}_j\right)\widehat{\Pr}^{\ast b}_{\mathbb{P}}\left(S = S^{\ast b}_j \mid A = a''\right)}{\sum_{a}\widehat{\Pr}^{\ast b}_{\mathbb{P}}\left(A = a \mid G = G^{\ast b}_j\right)\widehat{\Pr}^{\ast b}_{\mathbb{P}}\left(S = S^{\ast b}_j \mid A = a\right)}
    \end{align*}
    }
    Compute
    \[
    \hat{\delta}^{\ast b} \equiv \frac{\sum_{j = 1}^{n_{\mathcal{P}}}\widehat{\Pr}^{\ast b}_{\mathbb{P}}\left(A = a' \mid S = S^{\ast b}_j, G = G^{\ast b}_j\right)Y^{\ast b}_j}{\sum_{j = 1}^{n_{\mathcal{P}}}\widehat{\Pr}^{\ast b}_{\mathbb{P}}\left(A = a' \mid S = S^{\ast b}_j, G = G^{\ast b}_j\right)} - \frac{\sum_{j = 1}^{n_{\mathcal{P}}}\widehat{\Pr}^{\ast b}_{\mathbb{P}}\left(A = a'' \mid S = S^{\ast b}_j, G = G^{\ast b}_j\right)Y^{\ast b}_j}{\sum_{j = 1}^{n_{\mathcal{P}}}\widehat{\Pr}^{\ast b}_{\mathbb{P}}\left(A = a'' \mid S = S^{\ast b}_j, G = G^{\ast b}_j\right)}
    \]
}
Output $(1-\alpha)$-level percentile bootstrap confidence interval 
\[
\left( \hat{\delta}_B^{(\alpha/2)}, \hat{\delta}_B^{(1-\alpha/2)} \right)
\]
where $\hat{\delta}_B^{(\alpha)}$ is the empirical $\alpha$-percentile of the $\hat{\delta}^{\ast b}$
\end{algorithm}

As mentioned in Section \ref{sec:related_work}, recent work has proposed an alternative approach: setting the sampling distribution to the posterior distribution obtained by updating an assumed prior with the ACS race-by-geolocation estimates \citep{derby2024}. We do not adopt this approach because, as discussed above, the Census Bureau offers its own account of the uncertainty of its estimates. This uncertainty is multifaceted---it includes considerations like the Census Bureau's sampling scheme and survey nonresponse, as well as adjustments the Census Bureau has made to account for them---and inscrutable to the general public. So, rather than impose our own model for this uncertainty by specifying an ultimately arbitrary prior, we prefer to use the model offered by the Census Bureau, which is best-positioned to develop one.

One potential advantage to assuming a prior distribution instead of using the Census Bureau's uncertainty model is that it can accommodate a superpopulation framework for the race probability model. For example, a researcher can assume an abstract superpopulation of which each year's demographic composition is a sample. The prior distribution characterizes the superpopulation of race probabilities and is updated by a given year's observed demographic composition. To our knowledge, the Census Bureau's uncertainty model cannot accommodate such a superpopulation framework because it only models uncertainty arising from its survey sampling procedure. In our view, however, this limitation is significant only if the superpopulation parameters of the race probability model are of independent interest. In most applications---like the study of racial disparities---the race probability model is merely nuisance.

Consider, for example, the task of estimating racial disparities in tax audit rates.\footnote{\citet{elzayn2023measuring} apply our proposed approach to this setting to quantify the uncertainty associated with their tax audit disparity estimates.} It is true that researchers might be interested in the racial disparity at the superpopulation level (e.g., as a parameter of an abstract data-generating process that produces the observed tax audit rates by race each year). But, to estimate such a disparity, they necessarily use observed tax audit data from certain, well-defined years. Suppose that the race of the taxpayer in each audit decision is unavailable, so the researchers impute it using BISG with ACS data from the relevant years. Quantifying the uncertainty of any individual race imputation and how it affects the uncertainty of the final disparity estimate requires only an understanding of the error of the year-specific BISG model used for that imputation; it does not require reference to any BISG (or other race probability) model at the superpopulation level.

\subsection{BISG Simulations}
\label{sec:bisg_sim}

We show via simulation that the uncertainty of BISG imputations generally has little effect on the variance of the resulting racial disparity estimate. For this simulation, we use the 2017-2021 ACS 5-year estimates of the racial composition of each census block group and the 2010 Census Bureau surname table. We use the following data-generating process for both the training and primary populations:
\begin{itemize}
    \item The proxy tuple $(G, S)$ is sampled i.i.d. from the marginal census block group and surname frequencies given by the ACS estimates and the Census Bureau surname table, excluding tuples where the racial composition of the surname is withheld by the Census Bureau and tuples where the surname and census block group have mutually exclusive racial compositions.
    \item The outcome $Y$ is i.i.d. standard normal, independent of $G$, $S$, and $A$: $Y \sim \mathcal{N}\left(0, 1\right)$.
\end{itemize}
Because the race probability model is estimated using BISG and the outcome is independent of race, no concrete race indicators need to be generated for this simulation. In each of 100 simulation repetitions, we draw 1,000 $(G, S, Y)$ tuples as the primary dataset $\mathcal{P}$. On this dataset, we estimate the average outcome $\hat{\delta}$ for each race using BISG-estimated probabilities. We then estimate the standard error using both the dual-bootstrap and the single-bootstrap.

\begin{figure}[h]
    \centering
    \includegraphics[width = 0.99\linewidth]{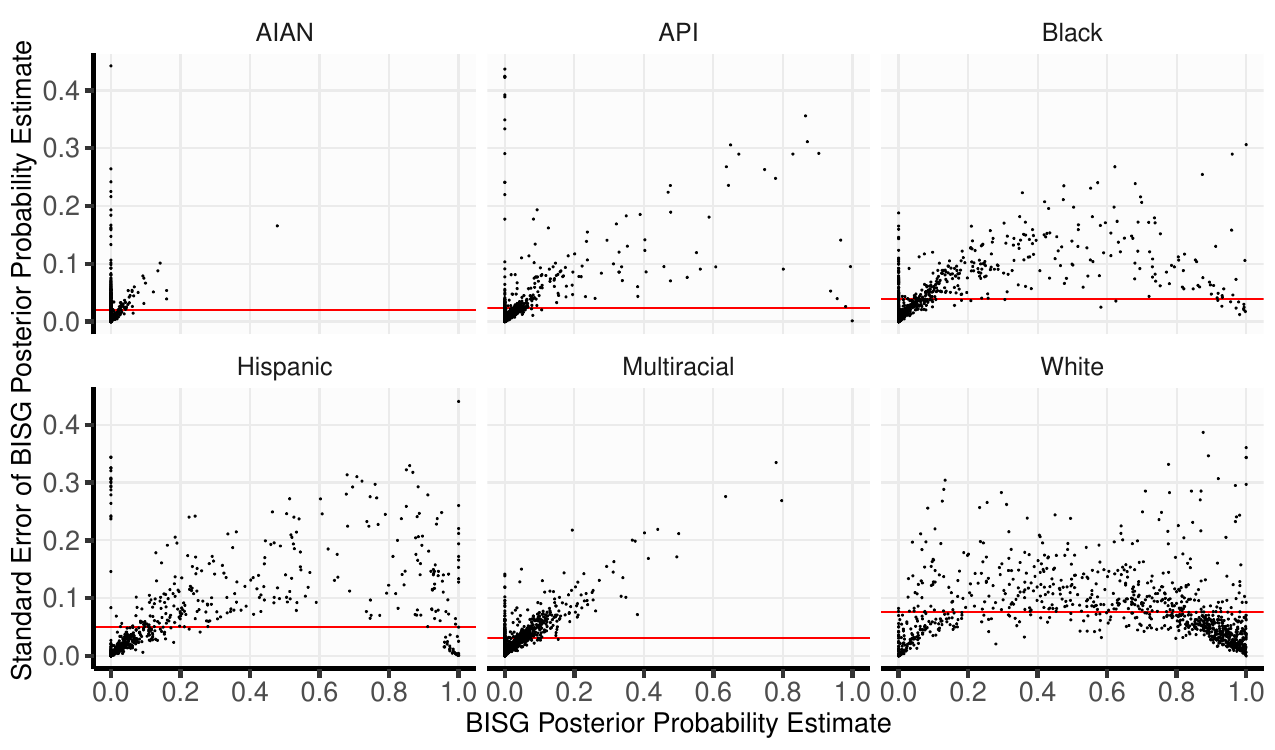}
    \caption{Bootstrap standard error of the BISG-estimated probability of being a given race plotted against the BISG-estimated probability in one simulation repetition. The horizontal red line indicates the average bootstrap standard error. ``AIAN'' is the abbreviation for American Indian and Alaska Native, and ``API'' is the abbreviation for Asian and Pacific Islander.}
    \label{fig:bisg_posteriors}
\end{figure}

Table \ref{tab:bisg_sim} reports the results. Overall, accounting for measurement uncertainty in this setting barely affects the resulting standard errors.\footnote{Consistent with this finding, \citet{elzayn2023measuring} obtain only slightly larger standard errors when they apply our method to estimate the uncertainty of their BISG-based estimates of tax audit disparities by race.} Figure \ref{fig:bisg_posteriors} offers one explanation: For most units, the bootstrap standard error of the posterior race probability is low, close to 0.05 on average. For comparison, the bootstrap standard errors of the race probabilities in the machine learning simulation of Section \ref{sec:ml_sim} are about 0.23, nearly five times larger. The standard errors are much smaller here likely because the BISG model is fairly rigid, and the ACS and surname prior probabilities that parameterize it are based on millions of individual-level training points. The notable exception in Table \ref{tab:bisg_sim} is the American Indian and Alaska Native group, for which the dual-bootstrap standard error is substantially less than the single-bootstrap standard error. As a theoretical matter, it might generally be possible for standard errors to decrease after properly accounting for measurement error; we do not prove so in our specific setting, but \citet{reifeis2022variance} show that this can occur in the closely related setting of IPW estimation of the average treatment effect on the treated. However, we believe that the specific reduction observed here is due to our handling of zero counts and our specific data-generating process, as described in more detail below.

\begin{table}
\centerline{
\begin{tabular}[t]{@{}lll@{}}
\toprule
& \multicolumn{2}{c}{Average Standard Error} \\
\cmidrule(l{3pt}r{3pt}){2-3} Race Group & Dual-Bootstrap & Single-Bootstrap\\
\midrule
American Indian and Alaska Native & 0.11 (0.02) & 0.18 (0.06)\\
Asian and Pacific Islander & 0.11 (0.01) & 0.11 (0.01)\\
Black & 0.07 (0.00) & 0.07 (0.00)\\
Hispanic & 0.06 (0.00) & 0.06 (0.00)\\
Multiracial & 0.07 (0.01) & 0.06 (0.01)\\
White & 0.04 (0.00) & 0.04 (0.00)\\
\bottomrule
\end{tabular}}
\caption{Average standard error of the estimated average outcome of each race, as estimated by the dual-bootstrap and the single-bootstrap when BISG is used for imputation. Standard deviations over the simulation repetitions are in parentheses. The dual-bootstrap's average estimate of the standard error is the same as that of the single-bootstrap except for the American Indian and Alaska Native group, for which it is lower. We offer one explanation for this in Figures \ref{fig:bisg_states}-\ref{fig:nm} and the associated discussion below. \label{tab:bisg_sim}}
\end{table}

\begin{figure}
    \centering
    \includegraphics[width = 0.99\linewidth]{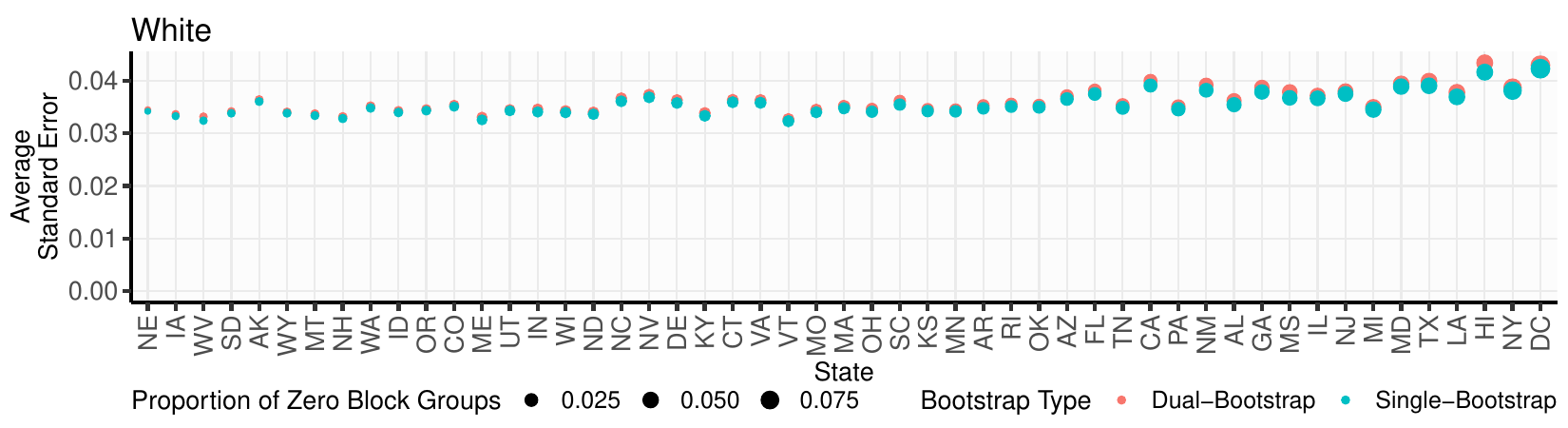}
     \includegraphics[width = 0.99\linewidth]{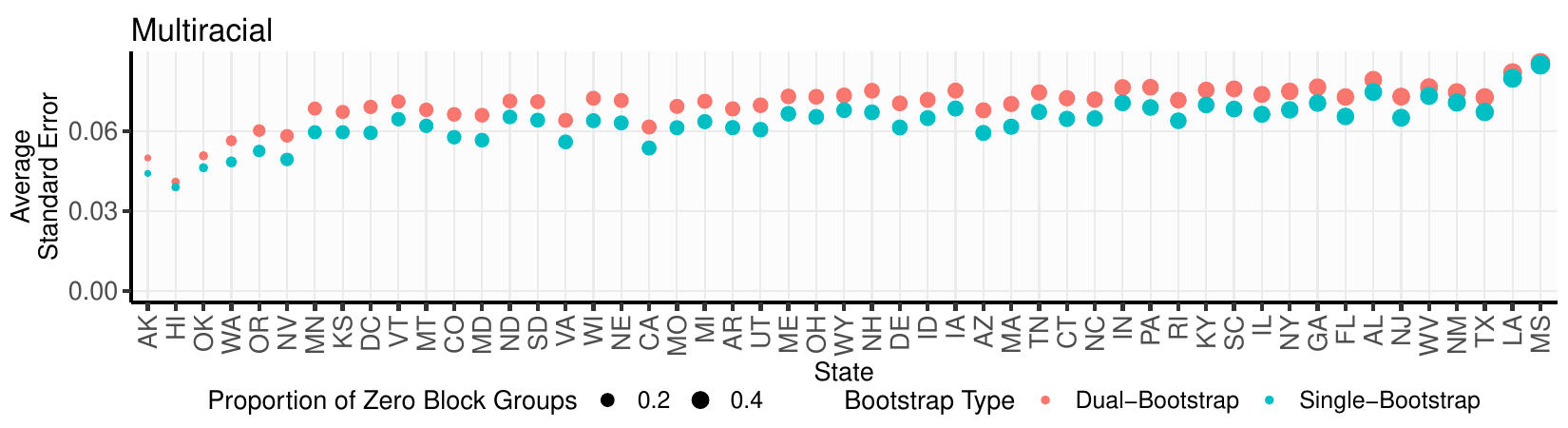}
    \includegraphics[width = 0.99\linewidth]{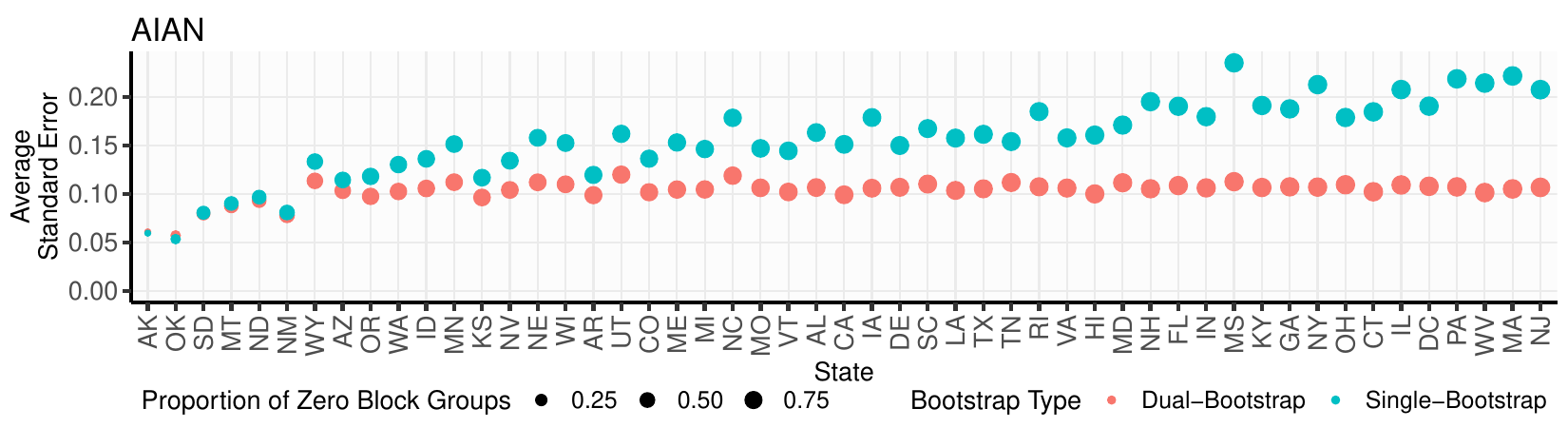}
    \caption{Dual-bootstrap and single-bootstrap standard errors of the estimated average outcome for the White, Multiracial, and American Indian and Alaska Native (AIAN) race groups in each state. The states are ordered by the proportion of census block groups in which the American Community Survey estimates there are zero people of the given race.}
    \label{fig:bisg_states}
\end{figure}

Although measurement uncertainty appears to be of nominal significance marginally over the entire population of the United States, we find evidence that it, and the way it is modeled, can be influential in certain situations. To illustrate, we rerun the above simulations for each state---that is, we sample census block groups from the marginal frequencies given by the ACS estimates within each state. Figure \ref{fig:bisg_states} shows the state-by-state results for three racial groups that we highlight here because of their particularly prominent trends; Appendix \ref{sec:additional_bisg_sims} contains corresponding figures for other racial groups. In most states, accounting for measurement uncertainty has essentially no effect on the uncertainty of the average outcome estimate for White people; it increases the uncertainty of the average outcome estimate for multiracial people; and it decreases the uncertainty of the average outcome estimate for American Indians and Alaska Natives.

We believe that the key to understanding these seemingly incompatible phenomena lies primarily in (1) the prevalence of each race group overall, (2) the geographic concentration of certain race groups, and (3) the distribution of the outcome among race groups. White people are much more prevalent than multiracial people overall: The 2017-2021 ACS data we use estimates that 59\% of the population is White, while 3\% are multiracial. Thus, ACS estimates of the proportion of White people in each census block group are more precise---in other words, have less measurement uncertainty---than estimates of the proportion of multiracial people. This explains why measurement uncertainty increases the uncertainty of the average outcome estimate for multiracial people more than it does for White people.

American Indians and Alaska Natives are even less prevalent than multiracial people overall: The 2017-2021 ACS data we use estimates that about 0.6\% of the population is American Indian or Alaska Native. But accounting for measurement uncertainty generally \textit{decreases} the uncertainty of the average outcome estimate in these simulations because they are a geographically concentrated minority: As Figure \ref{fig:bisg_states} shows, the ACS estimates that there are zero American Indians and Alaska Natives in most census block groups in most states. As discussed in Section \ref{sec:bisg_procedure}, we use the ACS estimated margin of error instead of the ACS variance replicates to approximate the sampling distribution of such estimates since it is possible that American Indians and Alaska Natives in fact reside in those block groups and were simply not sampled by ACS. Accounting for measurement uncertainty in this way gives nonzero weight to people who otherwise would have none. When the outcomes of these people are informative of the average outcome of American Indians and Alaska Natives---as they are in this simulation, since all units have outcomes drawn from a standard normal distribution---giving them nonzero weight increases the effective sample size and thus decreases the standard error of the average outcome estimate. This phenomenon likely explains the overall decrease in the standard error for American Indians and Alaska Natives reported in Table \ref{tab:bisg_sim}.

\begin{figure}[h]
    \centering
    \includegraphics[width = 0.99\linewidth]{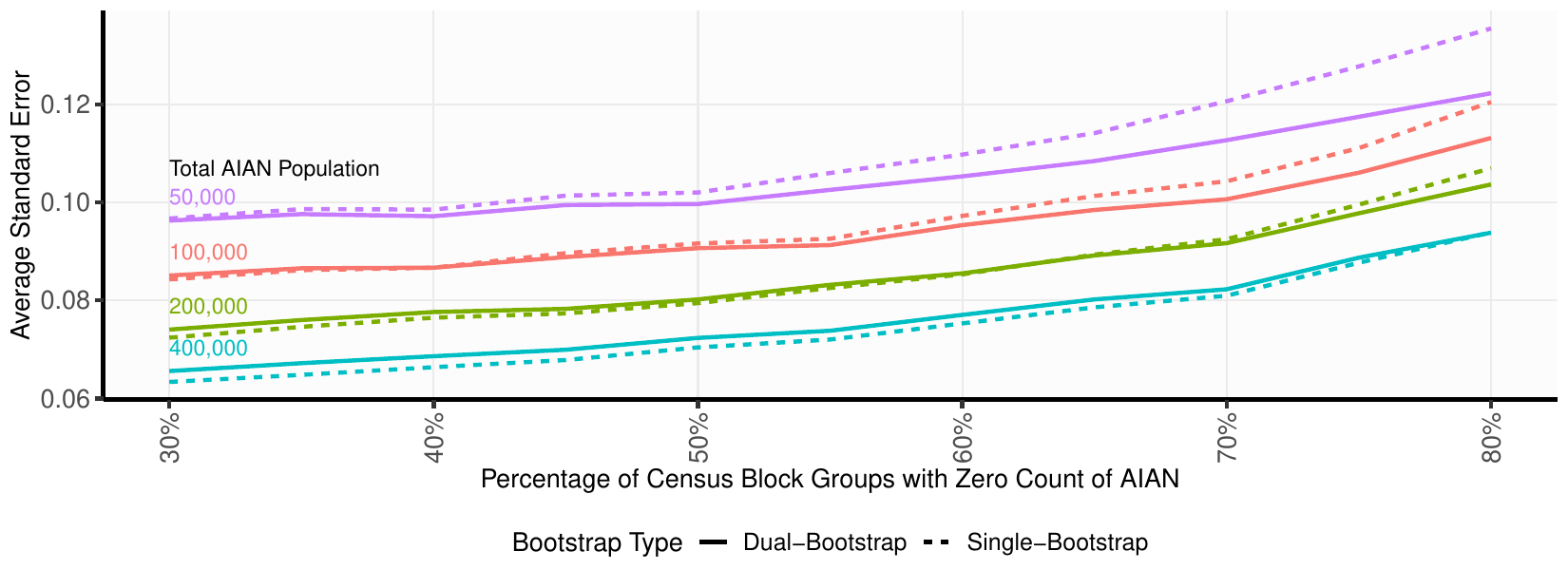}
    \caption{Dual-bootstrap and single-bootstrap standard errors of the estimated average outcome for the American Indian and Alaska Native (AIAN) race group in New Mexico when American Community Survey estimates of the total counts and geographic concentrations of AIAN people in the state are artificially altered.}
    \label{fig:nm}
\end{figure}

We illustrate some of these dynamics through an additional simulation focused on the American Indian and Alaska Native population in New Mexico. In this simulation, we generate synthetic states by taking ACS estimates from New Mexico and altering (1) the total prevalence of American Indian and Alaska Native people in the state and (2) the percentage of census block groups in the state in which zero American Indian and Alaska Native people are estimated to reside. Appendix \ref{sec:new_mexico_bisg_sims} describes this process in detail. We then rerun the previous simulation on each synthetic state. As Figure \ref{fig:nm} shows, accounting for measurement uncertainty decreases standard errors the less prevalent and the more concentrated the race group is.

These results for the American Indian and Alaska Native group demonstrate that a proper accounting of uncertainty in zero-count geolocations can, in specific circumstances, be influential. Our choice to take at face value the Census Bureau's margins of error when quantifying the uncertainty of the estimated average outcomes of a race group gives influence to people in geolocations where ACS estimates no people of that race group reside. This motivation to properly leverage data from zero-count geolocations also underlies some of the work of \citet{imai2022}---though they focus on how accounting for the migration of racial minorities since the last decennial census can improve imputation accuracy, whereas we focus on how accounting for the possible nonselection of racial minorities in ACS sampling can improve uncertainty quantification of downstream estimates. We emphasize that in practice, however, properly accounting for uncertainty in zero-count geolocations might not have as drastic or counterintuitive an effect as shown here, where the outcomes of all simulated units are equally informative for all race groups. On the contrary, it might in some cases increase standard errors if the outcomes of people in zero-count geolocations are substantially different from those of the target race group. In other cases, it might have no effect on balance.

Taken together, the phenomena identified in the simulations above highlight that, although the uncertainty of BISG imputations might not be substantial in studies of the general U.S. population, properly accounting for it in studies of particular geographic areas or demographic groups can be important to ensuring that the resulting inference is neither conservative nor anti-conservative. More generally, our finding is consistent with a broader literature on the challenges of race imputation for certain demographic groups \citep[e.g.,][]{imai2022}. 

\section{Application}
\label{sec:application}

We apply the dual-bootstrap to study racial disparities in health outcomes using the American Family Cohort (AFC) dataset \citep{AFC}. The dataset contains electronic health records from the primary care visits of patients in the United States. Relevant features for our purposes include patient geolocation; first name; surname; self-reported race, which are provided as mapped to White, Black, Hispanic, Asian and Pacific Islander, American Indian and Alaska Native, Multiracial, and Other; and indicators for the diagnosis of asthma, obesity, and diabetes at any point during the time period covered by the dataset. We downsample the data due to computational constraints by taking a stratified random sample of 100,000 patients with the same race proportions as the full dataset. Although we do not adopt them here, general steps can be taken in practice to improve the computational efficiency of the bootstrap \citep[e.g.,][]{blb}.

We preprocess the dataset as follows. First, we produce race proxies by converting categorical geolocation, first name, and surname data into numerical race probability estimates. Specifically, we convert the surnames into ``prior probability'' features by computing the probability of each of the six race categories (excluding Other) given surname based on the Census Bureau's 2010 surname table. And we convert the first names and geolocations into ``update'' features by computing the probability of the first name or geolocation given each of the six race categories. We use mean imputation for any missing geolocation, first name, or surname probabilities and include a binary missingness indicator for each as a separate feature. Second, we randomly split the data into a primary dataset of size 20,000 and a training dataset of size 80,000. We mask self-reported race in the primary dataset and health outcomes in the training dataset.

On the training dataset, we fit a random forest of patients’ races on the processed features defined above. Although these same features could be run through BISG to output race probabilities, we choose to use a random forest because \citet{cheng2023} find that it produces more accurate estimates in this dataset. We allow for slight tuning of the random forest hyperparameters. Specifically, we perform a grid search of the following hyperparameters using 5-fold cross-validation.

\begin{description}
    \item[Number of Trees:] 100.
    \item[Maximum Tree Splits:] 20, 50, 100.
    \item[Proportion of Features Per Split $(p = 21)$:] $\sqrt{p}$, $0.5p$, $0.75p$.
    \item[Minimum Number of Units to Initiate Split:] 10, 25, 100.
\end{description}

We then apply the random forest to the primary dataset to estimate patients' race probabilities. With those estimates in hand, we estimate the incidence rates of asthma, obesity, and diabetes for each race in the primary dataset. To compute confidence intervals for each estimate, we use both the single-bootstrap, which retains the original random forest model, and the dual-bootstrap, which refits a new random forest model on bootstrapped draws of the training dataset. We run 100 bootstrap iterations.

\begin{figure}
    \centering
    \includegraphics[width = 0.99\linewidth]{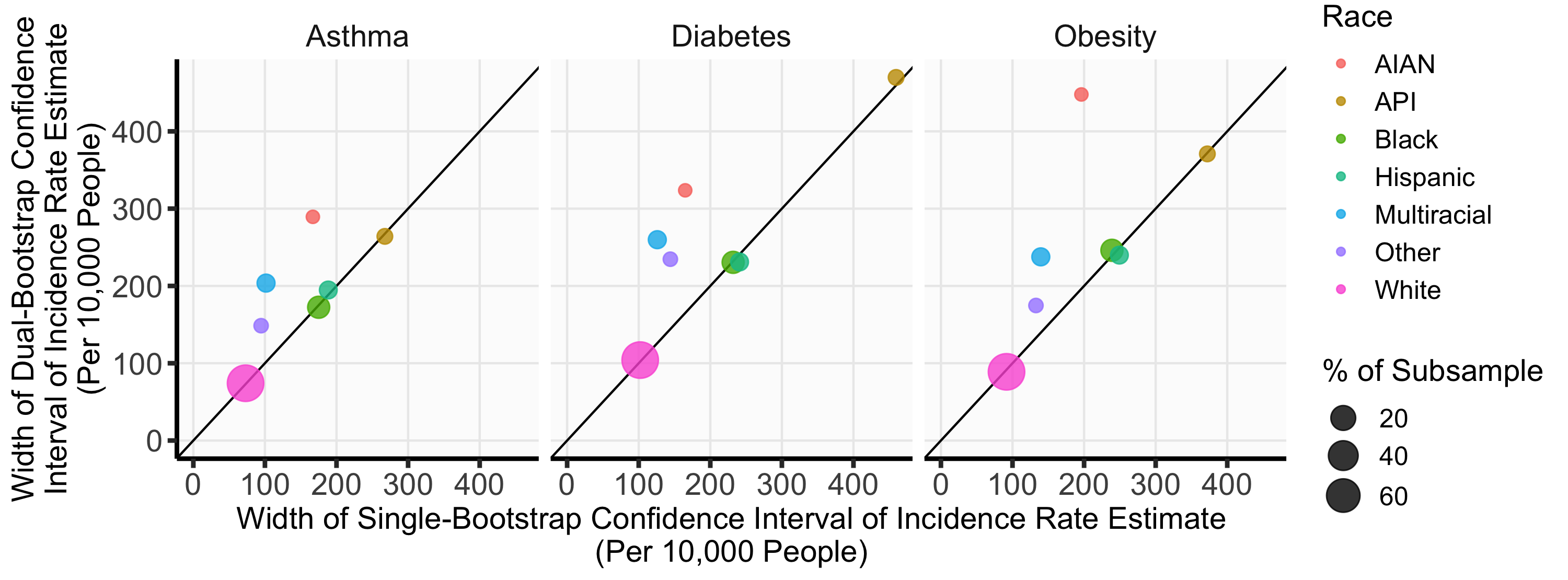}
    \caption{Widths of dual-bootstrap vs. single-bootstrap confidence intervals of the estimated prevalence of certain health conditions by race. This analysis was conducted on a 100,000-unit subsample of the American Family Cohort population with the same racial composition as the full population. Points are sized by the proportion of units in the subsample that are of the given race group. ``AIAN'' is the abbreviation for American Indian and Alaska Native, and ``API'' is the abbreviation for Asian and Pacific Islander.}
    \label{fig:afc_ses}
\end{figure}

\begin{figure}
    \centering
    \includegraphics[width = 0.99\linewidth]{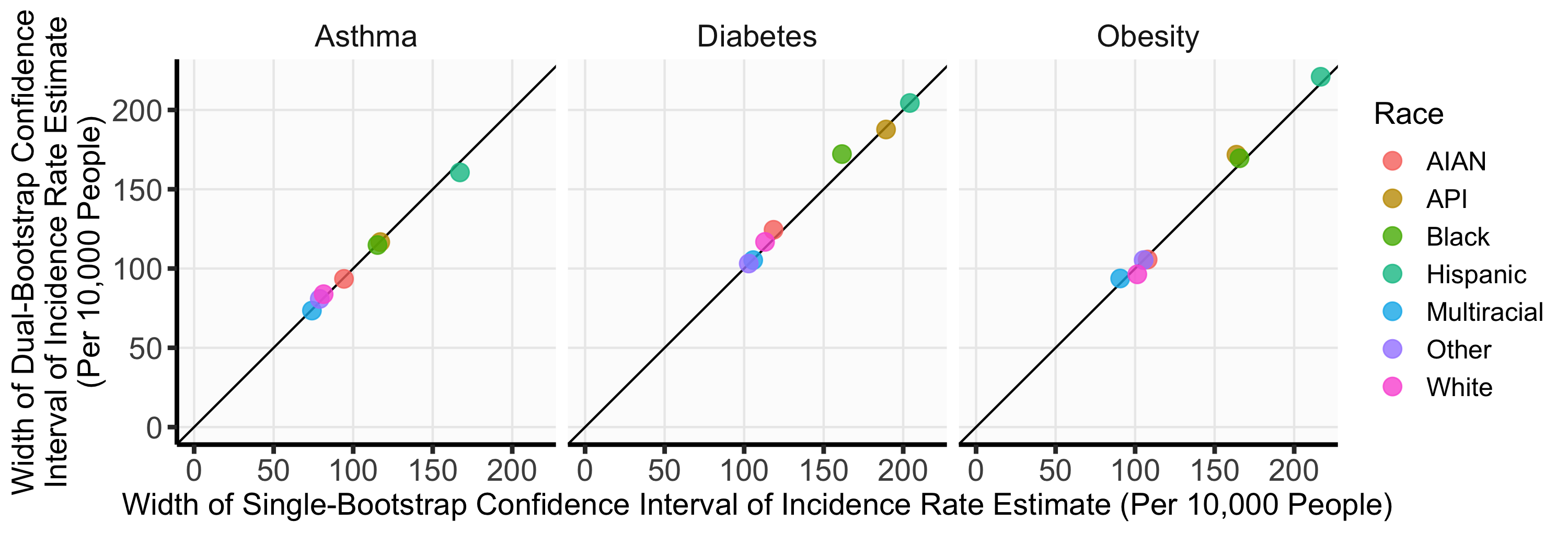}
    \caption{Widths of dual-bootstrap vs. single-bootstrap confidence intervals of the estimated prevalence of certain health conditions by race. This analysis was conducted on a 100,000-unit subsample of the American Family Cohort population where each race was equally represented in the training dataset. ``AIAN'' is the abbreviation for American Indian and Alaska Native, and ``API'' is the abbreviation for Asian and Pacific Islander.}
    \label{fig:afc_ses_equalprops}
\end{figure}

Figure \ref{fig:afc_ses} shows the results. In general, the single-bootstrap understates the uncertainties of the prevalence estimates compared to the dual-bootstrap. But this trend is not uniform across races. For Asian and Pacific Islander, Black, Hispanic, and White patients, the widths of the dual-bootstrap and single-bootstrap confidence intervals are essentially the same. On the other hand, the dual-bootstrap confidence intervals are substantially wider than the single-bootstrap confidence intervals for American Indian and Alaska Native, Multiracial, and Other patients---in some cases doubly so. This appears largely to be because those patients appear infrequently in the data on which the random forest model was trained, so their race probability estimates are more variable. As Figure \ref{fig:afc_ses_equalprops} shows, the dual-bootstrap confidence interval widths are nearly identical to the single-bootstrap ones when we alter our downsample so that all races are equally represented in the training dataset.
  
\section{Discussion}
\label{sec:discussion}

We propose a dual-bootstrap procedure to more accurately account for the uncertainty of race imputations that are subsequently used to estimate racial disparities and other race-specific outcomes. Our method is straightforward to implement, although complications can arise when the underlying data used to train the race probability model are unavailable. We offer one way of overcoming such difficulties in the specific case of BISG, an imputation model that is often parameterized by ACS race-by-geolocation estimates that are based on undisclosed microdata. Our simulation results suggest that the measurement uncertainty of BISG generally does not impact the uncertainty of downstream estimates, likely because it is a fairly rigid model with a relatively large sample size underpinning its parameter estimates. But it can be significant for specific race groups in specific geographies, with the potential to increase or decrease the standard error of downstream estimates, as our state-by-state results show. And we emphasize that despite its overall low variability---or perhaps because of it---BISG still suffers from bias, as others have shown and sought to improve \citep[e.g.,][]{imai2022}.

We see several opportunities for future work in this direction. Most immediately, an investigation of the theoretical properties of the dual-bootstrap when the race probability model falls outside the $Z$-estimator framework could be informative. On the practical side, a closer examination and improvement of some of the design choices made in our adaptation of the dual-bootstrap to BISG---such as our choice to use a normal distribution and other choices outlined in Appendix \ref{appendix:bisg}---could produce more accurate inference. Any changes or additions by the Census Bureau to the data products it publishes could help or hinder these efforts. We also see broader opportunities in this space. For example, the development of prospective heuristics for study design akin to a power analysis might prove useful to applied researchers. In some settings, researchers have a choice between analyzing a small dataset where race is observed and analyzing a larger dataset where race must be imputed. The need to account for measurement uncertainty---which, as shown in this paper, can be substantial or not---only complicates this choice. A set of heuristics that allows researchers to prospectively approximate the standard error that would result from each choice given certain parameters like the sample sizes of the datasets, the accuracy of the imputations, and the variability of the outcome might help with the decision.

\section*{Acknowledgements}
We thank Stanford's Center for Population Health Sciences for data, and Isabel Gallegos, Cameron Guage, Thomas Hertz, Peter Hull, Qiwei Lin, Mark Loewenstein, and participants in the NBER CRIW Race, Ethnicity, and Economic Statistics for the 21st Century conference for helpful comments. This material is based upon work supported by the National Science Foundation under Grant No. 1745640.

\newpage
\index{Bibliography@\emph{Bibliography}}%
\bibliographystyle{apalike}
\bibliography{cite.bib}

\newpage

\appendix
\section{Implementation Details for the BISG Adaptation of the Dual-Bootstrap}
\label{appendix:bisg}

We outline our approach to a few issues that can arise when implementing our dual-bootstrap adaptation to BISG.

\subsection{Zero Counts}
\label{sec:zero_counts}
In the ACS dataset, some geographic areas are estimated to have zero people of certain races. Intuitively, such ``zero counts'' are more common at the census block group level than at the ZIP code tabulation area level. In such cases, all 80 variance replicates also estimate zero people of that race, even though a full count may have revealed people of that race in that area. To nonetheless reflect this uncertainty, the Census Bureau reports a margin of error based on a different method from the successive differences replication method.

Whenever we encounter such instances of zero counts, we impute possible observed values for the 80 variance replicates from a discrete uniform distribution with minimum value 0 and maximum value determined by the margin of error. In particular, we derive the estimated variance from the reported margin of error using a formula prescribed by the Census Bureau. The variance, combined with the fact that observed counts can never be less than 0, allows us to derive the maximum possible value of the discrete uniform distribution. We sample from this distribution 80 times independently and replace the zero counts in the variance replicates with these values (and update the estimated total counts across all races in the variance replicates) for purposes of estimating the covariance matrix $\hat{\Sigma}$. 

Our choice of parametric distribution here has little downstream impact since it is used only to recover the variance, which maps directly to the margin of error given by the Census Bureau. But our estimates for zero-count races in a geolocation have approximately zero covariance with the estimates for other races even though this might not be the case in reality. We leave an examination of the significance of this choice and possible improvements to it to future work.

\subsection{Impermissible Sampled Probabilities}
\label{sec:impermissible}
In some cases, the draws of vectors $\widehat{\Pr}^{\ast b}_{\mathbb{P}}\left(A \mid G = g\right) \sim \mathcal{N}\left(\hat{\mu}_g, \hat{\Sigma}_g\right)$ will include elements that are less than 0 or greater than 1. This arises because we assume that the sampling distribution of the conditional race-by-geolocation probability estimates is multivariate normal. In such instances, we simply round the elements to 0 or 1 accordingly. The rounding to 1 is not strictly necessary, since the normalization that occurs in Bayes' Theorem implicitly handles it; but the rounding to 0 appears to be necessary.

This problem likely can be avoided by imposing an alternative form on the sampling distribution. For example, there might exist a unique set of parameters that best fit $\hat{\mu}_g$ and $\hat{\Sigma}_g$ as a Dirichlet distribution. If so, then modeling the sampling distribution as a Dirichlet with those parameters instead would sidestep this issue. We also note here that the densities of the Dirichlet distribution and the multivariate normal distribution with the same means and covariances converge asymptotically \citep{ouimet2022multivariate}. This might suggest that this problem is less significant in relatively large sample sizes like the ACS, but more research is needed to be sure.

\subsection{Mutually Exclusive Conditional Probabilities}
In some cases, the conditional surname-by-race probabilities and the conditional race-by-geolocation probabilities are incompatible. For example, an individual might have a surname that, according to the 2010 surname table, only White people have. But he or she might live in a census block group that, according to ACS estimates, has no White people. This problem is not unique to the dual-bootstrap; it can occur in any application of BISG. But it is more likely to occur when applying the dual-bootstrap, which calls for repeated computation of BISG probabilities on resamples of the training data.

Because this problem extends beyond the dual-bootstrap, we do not propose any particular solution. For purposes of the simulations in Section \ref{sec:bisg}, however, our stopgap approach is to give primacy to the surname probabilities: If the conditional surname-by-race probabilities and the conditional race-by-geolocation probabilities are incompatible, we simply do not update the former with the latter.

\newpage

\section{Additional Details on BISG Simulations}
\label{sec:additional_bisg}

\subsection{Additional State-by-State Results}
\label{sec:additional_bisg_sims}

\begin{figure}[h]
    \centering
    \includegraphics[width = 0.99\linewidth]{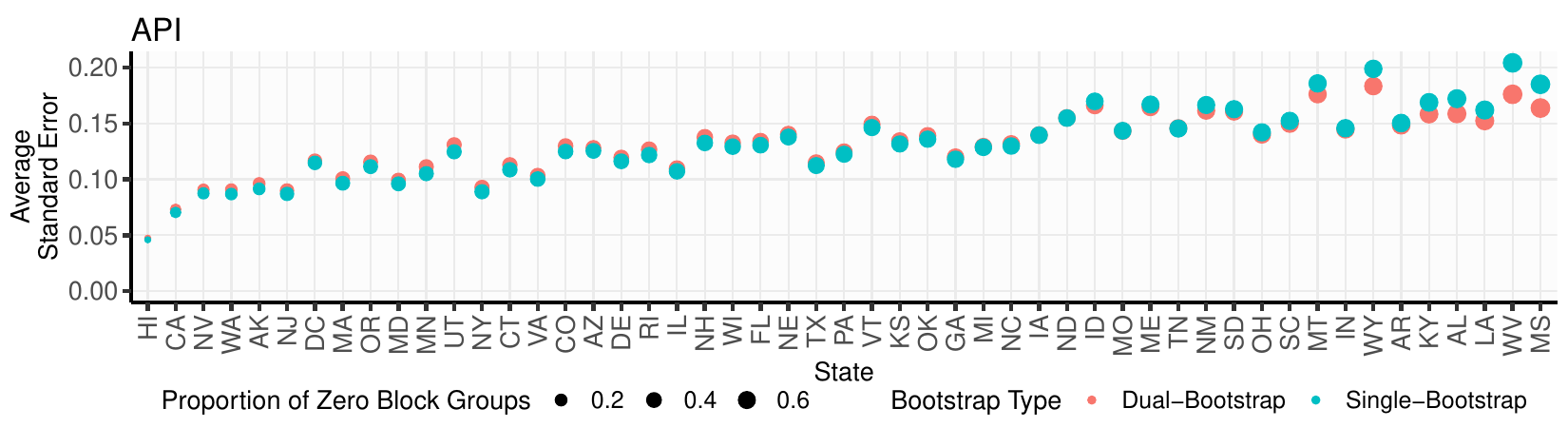}
    \includegraphics[width = 0.99\linewidth]{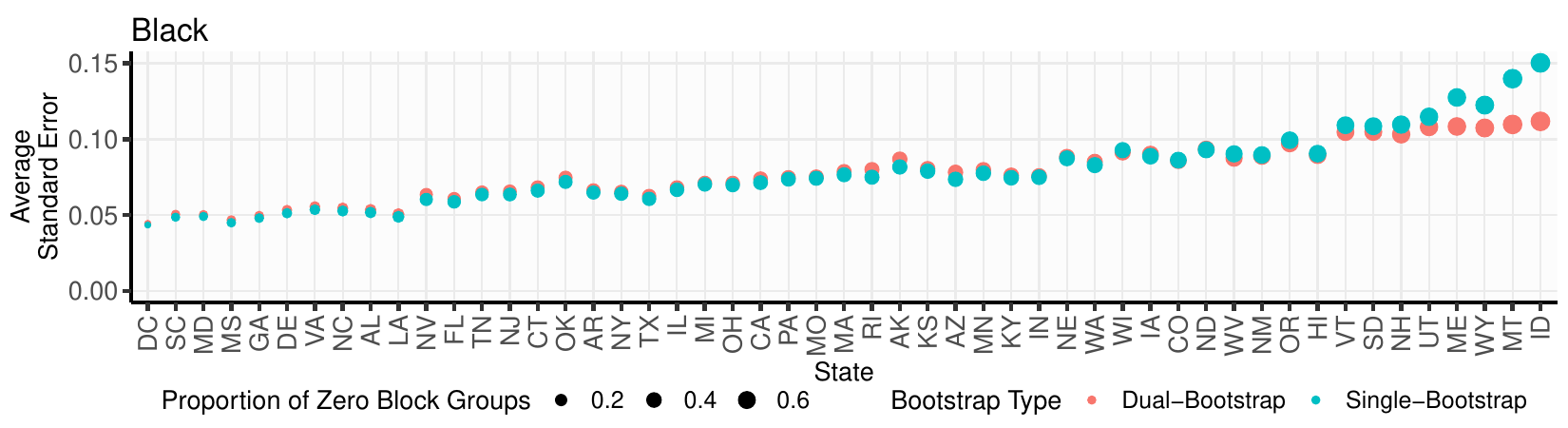}
    \includegraphics[width = 0.99\linewidth]{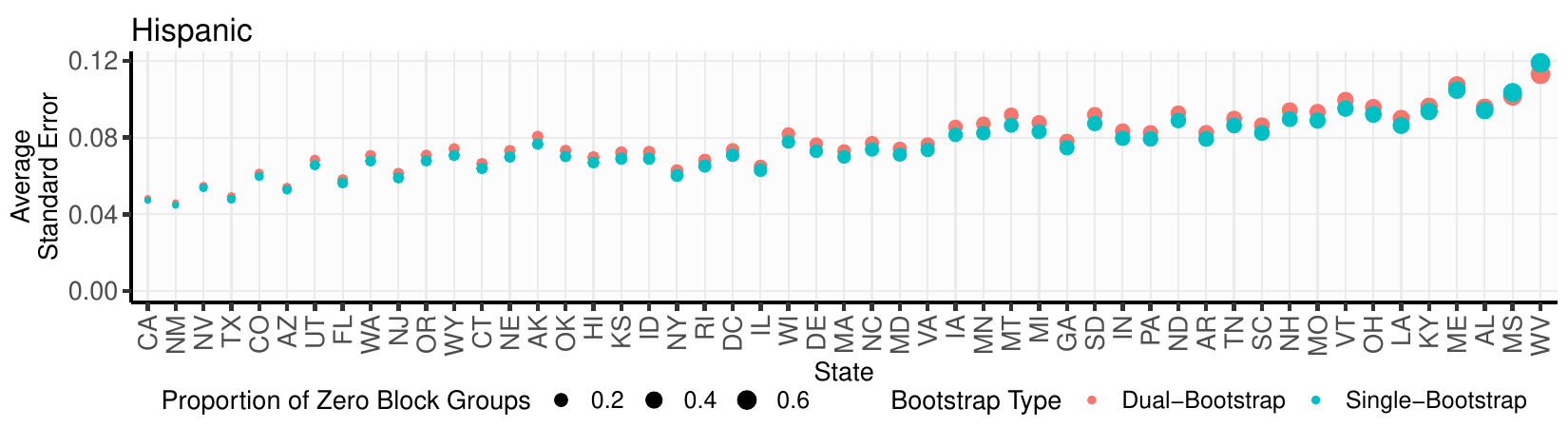}
    \caption{Dual-bootstrap and single-bootstrap standard errors of the estimated average outcome for the Asian and Pacific Islander (API), Black, and Hispanic race groups in each state. The states are ordered by the proportion of census block groups in which the American Community Survey estimates there are zero people of the given race.}
    \label{fig:bisg_states_appendix}
\end{figure}

Figure \ref{fig:bisg_states_appendix} shows the results of the state-by-state simulation in Section \ref{sec:bisg_sim} for the remaining race categories that we study. Although the trends shown here are less pronounced, we believe they can be interpreted within the framework described in Section \ref{sec:bisg_sim}.

\subsection{Additional Details on New Mexico Simulation}
\label{sec:new_mexico_bisg_sims}

In this section, we describe in more detail the New Mexico simulation reported in Section \ref{sec:bisg_sim}. We select New Mexico for illustrative purposes and focus on the effect of varying total size and census block group concentration of the American Indian and Alaska Native (AIAN) population in the state on the standard error of the average group outcome estimate. The following simulation can be conducted for any state and any race group. 

The simulation follows the same general procedure as the state-by-state simulation in Section \ref{sec:bisg_sim} for just New Mexico, except we modify the 2017-2021 ACS 5-year estimates of the AIAN composition of each census block group in the state. The actual estimated total population of AIAN in New Mexico is 181,021, and about 49\% of census block groups are estimated to have zero AIAN people. We vary the proportion of census block groups with zero-count AIAN from 30\% to 80\%, and also vary the total population of AIAN among the values 50,000, 100,000, 200,000, and 400,000. 

When modifying the proportion of census block groups with zero-count AIAN, we start with the existing distribution of AIAN counts in census block groups. We decrease the proportion of zero-count AIAN census block groups by randomly selecting zero-count AIAN census block groups (without replacement) and assigning them all the AIAN information (including ACS margins of error and variance replicate estimates) of randomly selected non-zero-count AIAN census block groups (with replacement). Similarly, we increase the proportion of zero-count AIAN census block groups by randomly selecting non-zero-count AIAN census block groups (without replacement) and assigning them the AIAN information of zero-count AIAN census block groups (with replacement). To then achieve the desired total population of AIAN, we scale all non-zero counts (and margins of error) of AIAN proportionally up or down. The result is that each synthetically generated New Mexico has different total sizes and census block group concentrations of AIAN, but a similarly shaped distribution of AIAN among non-zero-count AIAN census block groups.

Having modified the records of the ACS table that report AIAN information, we simply follow through with the rest of the procedure described in Section \ref{sec:bisg} and Appendix \ref{appendix:bisg}, including estimating the covariance matrix with zero-count adjustments, implementing Algorithm \ref{alg:bisg}, and conducting the standard error estimation simulation with 1,000 tuples in our synthetically generated New Mexico.

\newpage

\section{Asymptotic Normality of the Dual-Bootstrap}
\subsection{Proof of Asymptotic Normality for Logistic Regression}
\label{appendix:corrected_proof}

We prove that the dual-bootstrap produces asymptotically normal bootstrap statistics with properly calibrated variance under the simplifying assumption that the race probabilities obey the logistic regression model
\begin{equation}
\label{eq:logistic_model}
\Pr(A = 1 \mid Z) = \frac{\exp(\alpha^{\intercal}Z)}{1 + \exp(\alpha^{\intercal}Z)}.
\end{equation}
The proof can be readily extended to other race probability models that fall within the $Z$-estimation framework. The theorem may also hold for other race probability models as well, but we leave a proof of such results to future work.

For ease of notation, let $\mu_a \equiv \mathbb{E}\left[Y \mid A = a\right]$ for $a \in \{0, 1\}$.

Let

\begin{equation}
    \psi_\theta(z, a, y) \equiv
    \begingroup
\renewcommand*{\arraystretch}{2}
    \begin{bmatrix}
    \psi_{\alpha}(z, a, y)\\
    \psi_{1}(z, a, y)\\
    \psi_{0}(z, a, y)
    \end{bmatrix} \equiv
    \begin{bmatrix}
        z\left\{a - \frac{\strut\exp(\theta_{\alpha}^{\intercal}z)}{\strut1 + \exp(\theta_{\alpha}^{\intercal}z)}\right\}\\
        \frac{\strut\exp(\theta_{\alpha}^{\intercal}z)}{\strut1 + \exp(\theta_{\alpha}^{\intercal}z)}y - \frac{\strut\exp(\theta_{\alpha}^{\intercal}z)}{\strut1 + \exp(\theta_{\alpha}^{\intercal}z)}\theta_1\\
        \frac{\strut1}{\strut1 + \exp(\theta_{\alpha}^{\intercal}z)}y - \frac{\strut1}{\strut1 + \exp(\theta_{\alpha}^{\intercal}z)}\theta_0,
    \end{bmatrix}
\endgroup
\end{equation}
and assume that $\theta \equiv \begin{bmatrix} \theta_{\alpha} & \theta_1 & \theta_0\end{bmatrix}^{\intercal} \in \Theta \subset \mathbb{R}^p$ where $\Theta$ is open and $p < \infty$ is fixed.

Then, defining the map $\theta \mapsto \Psi(\theta) \equiv P\psi_{\theta}$, note that $\theta_0 \equiv \begin{bmatrix} \alpha & \mu_1 & \mu_0 \end{bmatrix}^{\intercal}$ satisfies $\Psi(\theta_0) = 0$. We show this coordinate by coordinate. First,
\begin{align}
    \mathbb{E}\left[Z\left\{A - \frac{\strut\exp(\alpha^{\intercal}Z)}{\strut1 + \exp(\alpha^{\intercal}Z)}\right\}\right] &= \mathbb{E}\left[ZA\right] - \mathbb{E}\left[Z\Pr(A = 1\mid Z)\right]\\
    &= \mathbb{E}\left[Z\Pr(A = 1 \mid Z)\right] - \mathbb{E}\left[Z\Pr(A = 1\mid Z)\right]\label{eq:tower}\\
    &= 0,
\end{align}
where \eqref{eq:tower} follows from the tower property conditioning on $Z$ and the fact that $A$ is binary. Second,
\begin{align}
    \mathbb{E}\left[\frac{\strut\exp(\alpha^{\intercal}Z)}{\strut1 + \exp(\alpha^{\intercal}Z)}Y - \frac{\strut\exp(\alpha^{\intercal}Z)}{\strut1 + \exp(\alpha^{\intercal}Z)}\mu_1\right] &= \mathbb{E}\left[\Pr(A = 1 \mid Z)Y\right] - \mathbb{E}\left[\Pr(A = 1 \mid Z)\right]\mu_1 \label{eq:mu1_id}\\
    &= \mathbb{E}\left[\Pr(A = 1 \mid Z)Y\right] - \mathbb{E}\left[\Pr(A = 1 \mid Z)\right]\frac{\mathbb{E}\left(AY\right)}{\Pr(A = 1)} \label{eq:total_exp}\\
    &= \mathbb{E}\left[\Pr(A = 1 \mid Z)Y\right] - \mathbb{E}\left(AY\right) \label{eq:total_exp2}\\
    &= \mathbb{E}\left[\Pr(A = 1 \mid Z)\mathbb{E}\left(Y \mid Z\right)\right] - \mathbb{E}\left\{\mathbb{E}\left(AY \mid Z\right)\right\} \label{eq:total_exp3}\\
    &= \mathbb{E}\left[\mathbb{E}(A \mid Z)\mathbb{E}\left(Y \mid Z\right) - \mathbb{E}\left(AY \mid Z\right)\right]\\
    &= 0, \label{eq:zero_cov}
\end{align}
where \eqref{eq:total_exp}, \eqref{eq:total_exp2}, and \eqref{eq:total_exp3} follow from the law of total expectation, and \eqref{eq:zero_cov} follows from our identifying assumption of zero covariance. The proof of the third coordinate is analogous.

Let $\hat{\theta}_n$ be an approximate zero of the estimating equation $\theta \mapsto \Psi_n(\theta) \equiv \mathbb{P}_n\psi_\theta$, and let $\hat{\theta}^\ast_n$ be an approximate zero of the bootstrapped estimating equation $\theta \mapsto \Psi^\ast_n(\theta) \equiv \mathbb{P}^\ast_n\psi_\theta$. By Theorem 10.16 of \citet{kosorok2008introduction},
\begin{equation}
    \sqrt{n}\left(\hat{\theta}_n - \theta_0\right) \overset{d}{\to} Z \sim \mathcal{N}\left(0, V^{-1}_{\theta_0}P\left[\psi_{\theta_0}\psi^{\intercal}_{\theta_0}\right]\left(V^{-1}_{\theta_0}\right)^{\intercal}\right)
\end{equation}
and
\begin{equation}
    \sqrt{n}\left(\hat{\theta}^{\ast}_n - \hat{\theta}_n\right) \underset{\ast}{\overset{P}{\rightsquigarrow}} k_0Z
\end{equation}
if five conditions hold. We verify each condition in turn. As a preliminary matter, note that Exercise 10.5.5 of \citet{kosorok2008introduction} already verifies each of the five conditions for the first coordinate of $\psi_\theta$. So we verify them for the remaining two coordinates, focusing without loss of generality on the first of the two.

\begin{description}
    \item[(A) For any sequence $\{\theta_n\} \in \Theta$, $\Psi(\theta_n) \to 0$ implies $\|\theta_n - \theta_0\| \to 0$.] \hphantom{a}
    \begin{proof}
        By assumption,
        \begin{equation}
            \mathbb{E}\left[\escore{\theta_{\alpha n}^{\intercal}Z}Y - \escore{\theta_{\alpha n}^{\intercal}Z}\theta_{1n}\right] \to 0.
        \end{equation}
        Distributing the expectation and dividing by both sides yields
        \begin{equation}
        \frac{\strut\mathbb{E}\left[\escore{\strut\theta_{\alpha n}^{\intercal}Z}Y\right]}{\strut\mathbb{E}\left[\escore{\strut\theta_{\alpha n}^{\intercal}Z}\right]} - \theta_{1n} \to 0,
        \end{equation}
        so it suffices to show that
        \begin{equation}
        \label{eq:show_converge}
        \frac{\strut\mathbb{E}\left[\escore{\strut\theta_{\alpha n}^{\intercal}Z}Y\right]}{\strut\mathbb{E}\left[\escore{\strut\theta_{\alpha n}^{\intercal}Z}\right]} \to \mu_1.
        \end{equation}
        Since we know from \eqref{eq:logistic_model} and \eqref{eq:mu1_id} that
        \begin{equation}
            \mu_1 = \frac{\strut\mathbb{E}\left[\escore{\strut\alpha^{\intercal}Z}Y\right]}{\strut\mathbb{E}\left[\escore{\strut\alpha^{\intercal}Z}\right]},
        \end{equation}
        it suffices to prove that the numerator and denominator of \eqref{eq:show_converge} each converge to their corresponding limit. We prove the denominator first. From Example 10.5.5 of \citet{kosorok2008introduction}, we can take as given that $\theta_{\alpha n} \to \alpha$. Since these are constants, this implies that $\theta_{\alpha n} \overset{p}{\to} \alpha$. Moreover, it is trivially true that an i.i.d. sequence $Z_1, Z_2, \ldots$ where each $Z_i$ is distributed as $Z$ satisfies $Z_n \overset{d}{\to} Z$. Then Slutsky's theorem implies that $\theta_{\alpha n} \pdot Z \overset{d}{\to} \alpha \pdot Z$. The continuous mapping theorem then implies that $\theta_{\alpha n}^{\intercal}Z \overset{d}{\to} \alpha^{\intercal}Z$. Since the logistic function is bounded and continuous, convergence in distribution implies that
        \begin{equation}
            \mathbb{E}\left[\escore{\strut\theta_{\alpha n}^{\intercal}Z}\right] \to \mathbb{E}\left[\escore{\strut\alpha^{\intercal}Z}\right].
        \end{equation}
        The proof for the numerator is similar but slightly more delicate. Again from Example 10.5.5 of \citet{kosorok2008introduction}, we can take as given that $\theta_{\alpha n} \to \alpha$. Since these are constants, this implies that $\theta_{\alpha n} \overset{p}{\to} \alpha$. Moreover, it is trivially true that an i.i.d. sequence $(Z_1, Y_1), (Z_2, Y_2), \ldots$ where each $(Z_i, Y_i)$ is distributed as $(Z, Y)$ satisfies $(Z_n, Y_n) \overset{d}{\to} (Z, Y)$. Since $\alpha$ is a constant, the portmanteau lemma implies that $(Z_n, Y_n, \theta_{\alpha n}) \overset{d}{\to} (Z, Y, \alpha)$. Then the continuous mapping theorem implies that
        \begin{equation}
        \escore{\strut\theta_{\alpha n}^{\intercal}Z}Y \overset{d}{\to} \escore{\strut\alpha^{\intercal}Z}Y.
        \end{equation}
        If $Y$ has a finite second moment, then
        \begin{equation}
        \mathbb{E}\left[\left|\escore{\strut\theta_{\alpha n}^{\intercal}Z}Y\right|^2\right] < \mathbb{E}\left[\left|Y\right|^2\right] < \infty 
        \end{equation}
        for all $n \in \mathbb{N}$, so
        \begin{equation}
            \escore{\strut\theta_{\alpha n}^{\intercal}Z}Y
        \end{equation}
        is uniformly integrable.\footnote{https://www.randomservices.org/random/expect/Uniform.html} This, combined with convergence in distribution, implies that
        \begin{equation}
            \mathbb{E}\left[\escore{\strut\theta_{\alpha n}^{\intercal}Z}Y\right] \to \mathbb{E}\left[\escore{\strut\alpha^{\intercal}Z}Y\right].
        \end{equation}
        Thus, the proof is complete.
    \end{proof}

    \item[(B) The class $\{\psi_\theta : \theta \in \Theta\}$ is strong Glivenko-Cantelli.] \hphantom{a}
    \begin{proof}
        \hphantom{a}As indicated by \citet{van2000asymptotic}, it suffices to show separately that each coordinate class is strong Glivenko-Cantelli. This can be done under several different regularity conditions. We assume two regularity conditions. First, we assume that each coordinate of $(Z, A, Y)$ is bounded almost surely---i.e., that $(Z, A, Y) \sim P$ where $P$ has measure zero outside a bounded subset of $\mathbb{R}^{p + 2}$. Second, we assume that $\Theta$ is bounded. Then let $\mathbb{R}^{p + 2} = \cup_j I_j$ be a partition in cubes of volume 1. Since each $\psi_1$ in the class has partial derivatives up to order $\alpha > (p + 2) / 2$ that are bounded by constants $M_j$ on each of the cubes $I_j$, Example 19.9 of \citet{van2000asymptotic} guarantees that, for any $V \geq (p + 2) / \alpha$,
        \begin{equation}
            \log N_{[]}\left(\epsilon, \{\psi_\theta : \theta \in \Theta\}, L_2(P)\right) \leq K\left(\frac{1}{\epsilon}\right)^V\left(\sum_{j = 1}^{\infty}\left(M_j^2P(I_j)\right)^{\frac{V}{V + 2}}\right)^{\frac{V +2}{2}}.
        \end{equation}
        Since $P$ has measure zero outside a bounded subset of $\mathbb{R}^{p + 2}$, the series converges for any $V \geq (p + 2) / \alpha$. Then, setting $V \in [(p + 2) / \alpha, 2)$, we see that the function class has finite bracketing integral:
        \begin{align}
            J_{[]}\left(1, \{\psi_\theta : \theta \in \Theta\}, L_2(P)\right) &\equiv \int_{0}^{1}\sqrt{\log N_{[]}\left(\epsilon, \{\psi_\theta : \theta \in \Theta\}, L_2(P)\right)}\,d\epsilon\\
            &\leq \int_{0}^{1}\sqrt{K\left(\frac{1}{\epsilon}\right)^V\left(\sum_{j = 1}^{\infty}\left(M_j^2P(I_j)\right)^{\frac{V}{V + 2}}\right)^{\frac{V +2}{2}}}\,d\epsilon\\
            &= \sqrt{K\left(\sum_{j = 1}^{\infty}\left(M_j^2P(I_j)\right)^{\frac{V}{V + 2}}\right)^{\frac{V +2}{2}}}\int_{0}^{1}\sqrt{\left(\frac{1}{\epsilon}\right)^V}\,d\epsilon\\
            &\leq C\int_{0}^{1}\left(\frac{1}{\epsilon}\right)^{\frac{V}{2}}\,d\epsilon\\
            &< \infty,
        \end{align}
        where $C$ is a constant.
        Thus, by Theorem 19.5 of \citet{van2000asymptotic}, the function class is Donsker. Hence, it is strong Glivenko-Cantelli \citep{kosorok2008introduction}. Note that we can probably relax the assumption that $P$ has measure zero if we instead assume it has a certain concentration.
    \end{proof}
    \item[{\parbox[t]{15cm}{(C) For some $\eta > 0$, the class $\mathcal{F} \equiv \left\{\psi_\theta:\theta \in \Theta, \left\|\theta - \theta_0\right\| \leq \eta\right\}$ is Donsker and $P(\psi_\theta - \psi_{\theta_0})^2 \to 0$ as $\|\theta - \theta_0\| \to 0$.}}]\hphantom{a}
    \begin{proof}
        The first statement follows immediately from our proof of (B). The second statement follows from similar logic to the proof of (A), assuming that $\mathbb{E}\left[\left|Y\right|^4\right] < \infty$. Observe that
        \begin{equation}
            P(\psi_\theta - \psi_{\theta_0})^2 = \mathbb{E}\left[\left\{\escore{\theta_{\alpha}^{\intercal}Z}(Y - \theta_1) - \escore{\alpha^{\intercal}Z}(Y - \mu_1)\right\}^2\right].
        \end{equation}
        Consider the first outer term obtained by squaring the inside:
        \begin{align}
        &\mathbb{E}\left[\left\{\escore{\theta_{\alpha}^{\intercal}Z}(Y - \theta_1)\right\}^2\right]\\
        &= \mathbb{E}\left[\left\{\escore{\theta_{\alpha}^{\intercal}Z}\right\}^2Y^2\right] - 2\theta_1\mathbb{E}\left[\left\{\escore{\theta_{\alpha}^{\intercal}Z}\right\}^2Y\right] + \theta_1^2\mathbb{E}\left[\left\{\escore{\theta_{\alpha}^{\intercal}Z}\right\}^2\right]\\
        &\to \mathbb{E}\left[\left\{\escore{\alpha^{\intercal}Z}\right\}^2Y^2\right] - 2\mu_1\mathbb{E}\left[\left\{\escore{\alpha^{\intercal}Z}\right\}^2Y\right] + \mu_1^2\mathbb{E}\left[\left\{\escore{\alpha^{\intercal}Z}\right\}^2\right],
        \end{align}
        where convergence occurs by application of the continuous mapping theorem and uniform integrability (again, assuming that the fourth moment of $Y$ is finite) to the fact that $(\theta_{\alpha}, \theta_1) \to (\alpha, \mu_1)$. Since this just is the second outer term obtained by squaring the inside, it suffices to show that the inner term converges to twice it. Observe that
        \begin{align}
            \mathbb{E}&\left[\left\{\escore{\theta_{\alpha}^{\intercal}Z}(Y - \theta_1)\right\}\left\{\escore{\alpha^{\intercal}Z}(Y - \mu_1)\right\}\right]\\ &= \mathbb{E}\left[\escore{\theta_{\alpha}^{\intercal}Z}\escore{\alpha^{\intercal}Z}\left\{Y^2 - \theta_1Y - \mu_1Y + \theta_1\mu_1\right\}\right].
        \end{align}
        Distributing and taking each term in turn, we have, by repeated application of the continuous mapping theorem,
        \begin{align}            \mathbb{E}\left[\escore{\theta_{\alpha}^{\intercal}Z}\escore{\alpha^{\intercal}Z}Y^2\right] &\to \mathbb{E}\left[\left\{\escore{\alpha^{\intercal}Z}\right\}^2Y^2\right]\\
        \theta_1\mathbb{E}\left[\escore{\theta_{\alpha}^{\intercal}Z}\escore{\alpha^{\intercal}Z}Y\right] &\to \mu_1\mathbb{E}\left[\left\{\escore{\alpha^{\intercal}Z}\right\}^2Y\right]\\
        \mu_1\mathbb{E}\left[\escore{\theta_{\alpha}^{\intercal}Z}\escore{\alpha^{\intercal}Z}Y\right] &\to \mu_1\mathbb{E}\left[\left\{\escore{\alpha^{\intercal}Z}\right\}^2Y\right]\\
        \theta_1\mu_1\mathbb{E}\left[\escore{\theta_{\alpha}^{\intercal}Z}\escore{\alpha^{\intercal}Z}\right] &\to \mu_1^2\mathbb{E}\left[\left\{\escore{\alpha^{\intercal}Z}\right\}^2\right].
        \end{align}
        Combining terms completes the proof.
    \end{proof}

    \item[(D) $P\|\psi_{\theta_0}\|^2 < \infty$ and $\Psi(\theta)$ is differentiable at $\theta_0$ with nonsingular derivative matrix $V_{\theta_0}$.]\hphantom{a}
    \begin{proof}
    The first part holds for $\psi_1$ and $\psi_0$ if $Y$ has finite second moment. To verify the second part, observe that
    \begin{align}
        \frac{\partial}{\partial \theta_{\alpha}}\psi_1 &= \frac{\exp(\theta_{\alpha}^{\intercal}z)}{\left(1 + \exp(\theta_{\alpha}^{\intercal}z)\right)^2}\left(y - \theta_1\right)z,\\
        \frac{\partial}{\partial \theta_{1}}\psi_1 &= \escore{\theta_{\alpha}^{\intercal}z}.
    \end{align}
    These continuous partial derivatives are uniformly bounded within a neighborhood of $\theta_0$:
    \begin{align}
        \left|\frac{\partial}{\partial \theta_{\alpha}}\psi_1\right| &\leq \left(\left|y\right| + \left|\mu_1\right| + \epsilon\right)\left|z\right|,\\
        \left|\frac{\partial}{\partial \theta_{1}}\psi_1\right| &\leq 1.
    \end{align}
    Moreover, these upper bounds are integrable if we assume that $\mathbb{E}\left[\left|Y\right|\right] < \infty$, $\mathbb{E}\left[\left|Z\right|\right] < \infty$, and $Z$ and $Y$ have finite variances. Thus, the Leibniz integral rule (applying the dominated convergence theorem and mean value theorem) implies that
    \begin{align}
        \frac{\partial}{\partial \theta_{\alpha}}\mathbb{E}\left[\psi_1\right]\Bigr|_{\theta = \theta_0} &= \mathbb{E}\left[\frac{\partial}{\partial \theta_{\alpha}}\psi_1\right]\Bigr|_{\theta = \theta_0} = \mathbb{E}\left[\frac{\exp(\alpha^{\intercal}Z)}{\left(1 + \exp(\alpha^{\intercal}Z)\right)^2}\left(Y - \mu_1\right)Z\right],\\
        \frac{\partial}{\partial \theta_{1}}\mathbb{E}\left[\psi_1\right]\Bigr|_{\theta = \theta_0} &= \mathbb{E}\left[\frac{\partial}{\partial \theta_1}\psi_1\right]\Bigr|_{\theta = \theta_0} = \mathbb{E}\left[\escore{\alpha^{\intercal}Z}\right].
    \end{align}
    A similar argument holds for the partial derivatives of $\psi_0$. Notably, one regularity condition for the derivative matrix to be nonsingular is that the expected value of the race probabilities must be bounded away from 0 and 1.
    \end{proof}

    \item[(E) $\Psi_n(\hat{\theta}_n) = o_P(n^{-1/2})$ and $\Psi^{\circ}_n(\hat{\theta}^{\circ}_n) = o_P(n^{-1/2})$.]\hphantom{a}
    \begin{proof}
        This follows for the last two coordinates of $\Psi_n(\hat{\theta}_n)$ because $\hat{\theta}_{1n}$ and $\hat{\theta}_{0n}$ are exact zeros of the estimating equation. The same is true of the last two coordinates of $\Psi^{\circ}_n(\hat{\theta}^{\circ}_n)$.
    \end{proof}
\end{description}

\subsection{Extending the $Z$-Estimator Framework to Other Models}\label{appendix:other_z_est}
In this section, we briefly describe how the linear disparity estimator that \citet{elzayn2023measuring} consider might fit into the $Z$-estimator framework used to prove asymptotic normality of the dual-bootstrap in Appendix \ref{appendix:corrected_proof}. The linear disparity estimator $\hat{\delta}_l$ is given by the estimated slope coefficient in the linear regression of $Y$ on the estimated probability $\widehat{\Pr}(A=1|Z)$ plus an intercept term.

We can formulate $\hat{\delta}_l$ as a $Z$-estimator. Specifically, let
\begin{equation}
    \psi_\theta(z, a, y) \equiv
    \begingroup
\renewcommand*{\arraystretch}{2}
    \begin{bmatrix}
    \psi_{\alpha}(z, a, y)\\
    \psi_{l}(z, a, y)
    \end{bmatrix} \equiv
    \begin{bmatrix}
        z\left\{a - \frac{\strut\exp(\theta_{\alpha}^{\intercal}z)}{\strut1 + \exp(\theta_{\alpha}^{\intercal}z)}\right\}\\
        \frac{\strut\exp(\theta_{\alpha}^{\intercal}z)}{\strut1 + \exp(\theta_{\alpha}^{\intercal}z)}\left(y - \frac{\strut\exp(\theta_{\alpha}^{\intercal}z)}{\strut1 + \exp(\theta_{\alpha}^{\intercal}z)}\theta_l \right)
    \end{bmatrix}
\endgroup
\end{equation}
and assume that $\theta \equiv \begin{bmatrix} \theta_{\alpha} & \theta_l \end{bmatrix}^{\intercal} \in \Theta \subset \mathbb{R}^p$ where $\Theta$ is open and $p < \infty$ is fixed. Then, defining the map $\theta \mapsto \Psi(\theta) \equiv P\psi_{\theta}$, we can show that $\theta_0 \equiv \begin{bmatrix} \alpha & \delta_l \end{bmatrix}^{\intercal}$ satisfies $\Psi(\theta_0) = 0$, where $\delta_l$ is the true disparity. If the same five conditions discussed in Appendix \ref{appendix:corrected_proof} also hold here, then the dual-bootstrap is asymptotically normal for the linear disparity estimator as well. We leave verification of these conditions to future work.
\end{document}